\documentclass[aps,a4paper,superscriptaddress,showpacs,preprintnumbers,amsmath,amssymb]{revtex4}

\usepackage{ulem}

\usepackage{psfrag} \usepackage{graphicx} \usepackage{dcolumn}
\usepackage{color} \usepackage{latexsym,amsfonts} \usepackage{bm}
\usepackage{amssymb}
\begin{document}

\title{Corrections of order $O(E^2_e/m^2_N)$, \\ caused by weak
  magnetism and proton recoil, \\to the neutron lifetime and
  correlation coefficients of the neutron beta decay}

\author{A. N. Ivanov}\email{ivanov@kph.tuwien.ac.at}
\affiliation{Atominstitut, Technische Universit\"at Wien, Stadionallee
  2, A-1020 Wien, Austria}
\author{R. H\"ollwieser}\email{roman.hoellwieser@gmail.com}
\affiliation{Atominstitut, Technische Universit\"at Wien, Stadionallee
  2, A-1020 Wien, Austria}\affiliation{Department of Physics,
  Bergische Universit\"at Wuppertal, Gaussstr. 20, D-42119 Wuppertal,
  Germany} \author{N. I. Troitskaya}\email{natroitskaya@yandex.ru}
\affiliation{Atominstitut, Technische Universit\"at Wien, Stadionallee
  2, A-1020 Wien, Austria}
\author{M. Wellenzohn}\email{max.wellenzohn@gmail.com}
\affiliation{Atominstitut, Technische Universit\"at Wien, Stadionallee
  2, A-1020 Wien, Austria} \affiliation{FH Campus Wien, University of
  Applied Sciences, Favoritenstra\ss e 226, 1100 Wien, Austria}
\author{Ya. A. Berdnikov}\email{berdnikov@spbstu.ru}\affiliation{Peter
  the Great St. Petersburg Polytechnic University, Polytechnicheskaya
  29, 195251, Russian Federation}

\date{\today}

\begin{abstract}
  We calculate the contributions of weak magnetism and proton recoil
  of order $O(E^2_e/m^2_N) \sim 10^{-5}$, i.e. to
  next-to-next-to-leading order in the large nucleon mass expansion,
  to the neutron lifetime and correlation coefficients of the neutron
  beta decay, where $E_e$ and $m_N$ are the electron energy and the
  nucleon mass, respectively.  We analyze the electron--energy and
  angular distribution for the neutron beta decay with a polarized
  neutron, a polarized electron and an unpolarized proton. Together
  with Wilkinson's corrections (Nucl. Phys. A {\bf 377}, 474 (1982)
  and radiative corrections of order $O(\alpha E_e/m_N) \sim 10^{-5}$
  (Phys. Rev. D {\bf 99}, 093006 (2019)), calculated as
  next--to--leading order corrections in the large nucleon mass $m_N$
  expansion to Sirlin's corrections of order $O(\alpha/\pi)$
  (Phys. Rev. {\bf 164}, 1767 (1967)), the corrections of order
  $O(E^2_e/m^2_N) \sim 10^{-5}$ provide an improved level of precision
  of the theoretical background of the neutron beta decay, calculated
  in the Standard Model, for experimental searches of contributions of
  interactions beyond the Standard Model.
\end{abstract}
 \maketitle

\section{Introduction}
\label{sec:introduction}

Nowadays experimental investigations of the neutron beta decay
\cite{Abele2008, Nico2009, Paul2009, Dubbers2011} with a polarized
neutron and unpolarized electron and proton are allowed at the level
of sensitivity of about $10^{-4}$ and even better \cite{Abele2016,
  Abele2018, Sirlin2018}. An analogous sensitivity is desirable also
for experimental investigations of the neutron beta decay with a
polarized neutron, a polarized electron and an unpolarized proton
\cite{Bodek2019}. The model--independent corrections of order of 
$10^{-3}$, caused by the radiative corrections of order
$O(\alpha/\pi)$ (so--called {\it outer} radiative corrections
\cite{Wilkinson1970}), where $\alpha$ is the fine--structure constant
\cite{PDG2020}, calculated to leading order in the large nucleon mass
$m_N$ expansion, together with corrections of order $O(E_e/m_N)$,
induced by weak magnetism and proton recoil, provide a robust
theoretical background for the experimental analysis of the neutron
beta decay at the level of $10^{-3}$. The notation ``large nucleon
mass $m_N$ expansion'' means that we make an expansion in powers of
$1/m_N$, were the nucleon mass is finite but much larger than the
momentum of the proton and energies and momenta of decay leptons. The
effect of the finite nucleon mass appears only to next-to-leading and
higher orders in the large nucleon mass $m_N$ expansion or in the
terms proportional to powers of $1/m_N$. For the neutron beta decay
with a polarized neutron and unpolarized electron and proton such a
theoretical background, including the radiative corrections of order
$O(\alpha/\pi)$ and the corrections $O(E_e/m_N)$, caused by weak
magnetism and proton recoil, were calculated in \cite{Sirlin1967,
  Shann1971, Sirlin1978, Bilenky1959, Wilkinson1982} (see also
\cite{Ando2004, Gudkov2006, Ivanov2013}). An universal {\it inner} 
radiative correction $\Delta^V_R$ \cite{Wilkinson1970} of order
$O(\alpha/\pi)$, which does not depend on the electron energy and
plays an important role for the correct theoretical description of the
neutron lifetime $\tau_n = 879.6(1.1)\,{\rm s}$ (see \cite{Ivanov2013}
and \cite{Abele2018, Sirlin2018}), was calculated in \cite{Sirlin1986,
  Sirlin2004, Sirlin2006, Seng2018, Seng2018a, Sirlin2019}. In turn,
for the neutron beta decay with a polarized neutron, a polarized
electron and an unpolarized proton the model--independent outer
radiative corrections of order $O(\alpha/\pi)$ and corrections of
order $O(E_e/m_N)$, caused by weak magnetism and proton recoil, were
calculated in \cite{Ivanov2017, Ivanov2018, Ivanov2019}.  In order to
promote an impetus for an improvement of experimental sensitivities
from the level of a few parts of $10^{-4}$ \cite{Paul2009, Abele2016,
  Abele2018} to the level of a few parts of $10^{-5}$ or even better
the neutron lifetime and the correlation coefficients of the neutron
beta decay should be calculated in the Standard Model (SM) at the
level of a few parts of $10^{-5}$. The step in this direction was done
by Wilkinson \cite{Wilkinson1982} (see also \cite{Ivanov2013,
  Ivanov2017, Ivanov2018, Ivanov2019, Severijns2018} for so--called
Wilkinson's corrections) and then in \cite{Ivanov2019a, Ivanov2020a},
where the radiative corrections of order $O(\alpha E_e/m_N) \sim
10^{-5}$ were calculated as next-to-leading order corrections in the
large nucleon mass expansion to Sirlin's  outer and inner 
radiative corrections of order $O(\alpha/\pi)$.

This paper is addressed to the calculation of the model--independent
corrections of order $O(E^2_e/m^2_N) \sim 10^{-5}$, caused by weak
magnetism and proton recoil. Together with Wilkinson's corrections
\cite{Wilkinson1982} (see also \cite{Ivanov2013, Ivanov2017,
  Ivanov2018, Ivanov2019, Severijns2018}) and radiative corrections of
order $O(\alpha E_e/m_N) \sim 10^{-5}$ \cite{Ivanov2019a,
  Ivanov2020a}, the corrections $O(E^2_e/m^2_N) \sim 10^{-5}$ should
provide an improved level of precision of the theoretical background
of the neutron beta decay, calculated in the SM. It is initiated to
promote experimental searches of contributions of interactions beyond
the SM \cite{ Bodek2019, Kozela2009, Kozela2012, Abele2019, Sun2020,
  Young2019} with experimental sensitivities of about a few parts of
$10^{-5}$.

The paper is organized as follows. In section \ref{sec:verteilung} we
give the electron--energy and angular distribution of the neutron beta
decay for a polarized neutron, a polarized electron and an unpolarized
proton with the account for the corrections of order
$O(E^2_e/m^2_N)$. We define the structure of the corrections of order
$O(E^2_e/m^2_N)$ to the neutron lifetime and to the correlation
coefficients.  In section \ref{sec:barX} we give the analytical
expressions for the corrections of order $O(E^2_e/m^2_N)$ to the
neutron lifetime and correlation coefficients. We give the corrections
$O(E^2_e/m^2_N)$ in the form of polynomials in the variable $E_e/E_0$,
where $E_0$ is the end--point energy of the electron--energy spectrum
\cite{Abele2008, Nico2009}. The coefficients of these polynomials are
calculated at the neglect of contributions of order of a few parts of
$10^{-6}$. In section \ref{sec:barX2} we give the total expressions
for the corrections $O(E^2_e/m^2_N)$ as polynomials in the variable
$E_e/E_0$ with a numerical analysis of the coefficients, where we have
kept the contributions of a few parts of $10^{-5}$ and $10^{-4}$ only.
Such a form of corrections under consideration makes them applicable
for the analysis of experimental data on asymmetries of the neutron
beta decay for the searches of contributions of interactions beyond
the SM at the level of $10^{-4}$ \cite{Abele2016, Ivanov2013,
  Ivanov2017, Ivanov2018, Ivanov2019} or even better at the level of a
few parts of $10^{-5}$ \cite{Ivanov2017, Ivanov2018, Ivanov2019,
  Gardner2013, Ivanov2020}. In section \ref{sec:Schluss} we discuss
the obtained results. In Appendix A we calculate the amplitude of the
neutron beta decay by taking into account the next-to-next-to-leading
order corrections $O(E^2_e/m^2_N)$, caused by weak magnetism and
proton recoil. In addition we take into account the contributions of
the isovector and axial--vector form factors of the nucleon defined in
the dipole approximation.  In Appendix B we give the contributions of
order $O(E_e/m_N)$ and $O(E^2_e/m^2_N)$ to the electron--energy and
angular distribution of the neutron beta decay with correlation
structures, which cannot be reduced to the correlation structure
proposed by Jackson {\it et al.}  \cite{Jackson1957}. In Appendix C we
adduce the corrections of order $O(E_e/m_N)$ to the neutron lifetime
and correlation coefficients, calculated in \cite{Ivanov2013,
  Ivanov2017, Ivanov2018, Ivanov2019}. We adduce these corrections for
the completeness of the analysis of the corrections of order
$O(E^2_e/m^2_N)$ to the neutron lifetime and the correlation
coefficients, since they give contributions of order of  $10^{-5}$ to the
corrections $O(E^2_e/m^2_N)$.  

\section{Electron--energy and angular distribution of the neutron
  beta decay with a polarized neutron, a polarized electron and an
  unpolarized proton}
\label{sec:verteilung}

For the calculation of the model--independent corrections of order
$O(E^2_e/m^2_N)$, induced by weak magnetism and proton recoil, to the
neutron lifetime and the correlation coefficients of the neutron beta
decay we use the standard effective Lagrangian of the $V-A$ weak
interaction with a real axial coupling constant $g_A$ and the
contribution of weak magnetism
\begin{eqnarray}\label{eq:1}
\hspace{-0.3in}{\cal L}_W(x) = -
\frac{G_F}{\sqrt{2}}\,V_{ud}\,\Big\{[\bar{\psi}_p(x)\gamma_{\mu}(1 -
  g_A \gamma^5)\psi_n(x)] + \frac{\kappa}{2 m_N}
\partial^{\nu}[\bar{\psi}_p(x)\sigma_{\mu\nu}\psi_n(x)]\Big\}
        [\bar{\psi}_e(x)\gamma^{\mu}(1 - \gamma^5)\psi_{\nu}(x)]
\end{eqnarray}
invariant under time reversal, where $G_F$ and $V_{ud}$ are the Fermi
weak coupling constant and the Cabibbo-Kobayashi-Maskawa (CKM) matrix
element \cite{PDG2020}, $\psi_p(x)$, $\psi_n(x)$, $\psi_e(x)$ and
$\psi_{\nu}(x)$ are the field operators of the proton, neutron,
electron and antineutrino, respectively, $\gamma^{\mu}$, $\gamma^5$
and $\sigma^{\mu\nu} = \frac{i}{2}(\gamma^{\mu}\gamma^{\nu} -
\gamma^{\nu}\gamma^{\mu})$ are the Dirac matrices \cite{Itzykson1980}
and $\kappa = \kappa_p - \kappa_n = 3.7059$ is the isovector anomalous
magnetic moment of the nucleon, defined by the anomalous magnetic
moments of the proton $\kappa_p = 1.7929$ and the neutron $\kappa_n =
- 1.9130$ and measured in nuclear magneton \cite{PDG2020}, and $m_N =
(m_n + m_p)/2 = 938.9188\, {\rm MeV}$ is the nucleon mass and $m_n =
939.5645\,{\rm MeV}$ and $m_p = 938.2721\,{\rm MeV}$ are the neutron
and proton masses \cite{PDG2020}, respectively. Then, $G_F$ and
$V_{ud}$ are the Fermi weak coupling constant and the
Cabibbo--Kobayashi--Maskawa (CKM) matrix element \cite{PDG2020}.

The electron--energy and angular distribution of the neutron beta
decays for a polarized neutron, a polarized electron and an
unpolarized proton was given by Jackson {\it et al.}
\cite{Jackson1957}. Following \cite{Gudkov2006, Ivanov2013,
  Ivanov2017, Ivanov2018, Ivanov2019} it can be written in the
following form (see also \cite{Ivanov2020} with a replacement $\lambda
= - g_A)$
\begin{eqnarray*}
\hspace{-0.15in}&&\frac{d^5 \lambda_n(E_e, \vec{k}_e,
  \vec{k}_{\bar{\nu}}, \vec{\xi}_n, \vec{\xi}_e)}{dE_e d\Omega_e
  d\Omega_{\bar{\nu}}} = (1 + 3
g^2_A)\,\frac{G^2_F|V_{ud}|^2}{32\pi^5}\,(E_0 - E_e)^2 \,\sqrt{E^2_e -
  m^2_e}\, E_e\,F(E_e, Z = 1)\,\zeta(E_e)\,\Big\{1 +
b(E_e)\,\frac{m_e}{E_e}\nonumber\\
\end{eqnarray*}
\begin{eqnarray}\label{eq:2}
\hspace{-0.15in}&& + a(E_e)\,\frac{\vec{k}_e\cdot
  \vec{k}_{\bar{\nu}}}{E_e E_{\bar{\nu}}} +
A(E_e)\,\frac{\vec{\xi}_n\cdot \vec{k}_e}{E_e} + B(E_e)\,
\frac{\vec{\xi}_n\cdot \vec{k}_{\bar{\nu}}}{E_{\bar{\nu}}} +
K_n(E_e)\,\frac{(\vec{\xi}_n\cdot \vec{k}_e)(\vec{k}_e\cdot
  \vec{k}_{\bar{\nu}})}{E^2_e E_{\bar{\nu}}}+
Q_n(E_e)\,\frac{(\vec{\xi}_n\cdot \vec{k}_{\bar{\nu}})(\vec{k}_e\cdot
  \vec{k}_{\bar{\nu}})}{ E_e E^2_{\bar{\nu}}}\nonumber\\
\hspace{-0.15in}&& + D(E_e)\,\frac{\vec{\xi}_n\cdot (\vec{k}_e\times
  \vec{k}_{\bar{\nu}})}{E_e E_{\bar{\nu}}} + G(E_e)\,\frac{\vec{\xi}_e
  \cdot \vec{k}_e}{E_e} + H(E_e)\,\frac{\vec{\xi}_e \cdot
  \vec{k}_{\bar{\nu}}}{E_{\bar{\nu}}} + N(E_e)\,\vec{\xi}_n\cdot
\vec{\xi}_e + Q_e(E_e)\,\frac{(\vec{\xi}_n\cdot \vec{k}_e)(
  \vec{k}_e\cdot \vec{\xi}_e)}{(E_e + m_e) E_e}\nonumber\\
\hspace{-0.15in}&& + K_e(E_e)\,\frac{(\vec{\xi}_e\cdot \vec{k}_e)(
  \vec{k}_e\cdot \vec{k}_{\bar{\nu}})}{(E_e + m_e)E_e E_{\bar{\nu}}} +
R(E_e)\,\frac{\vec{\xi}_n\cdot(\vec{k}_e \times \vec{\xi}_e)}{E_e} +
L(E_e)\,\frac{\vec{\xi}_e\cdot(\vec{k}_e \times
  \vec{k}_{\bar{\nu}})}{E_eE_{\bar{\nu}}}\Big\} + \frac{d^5
  \lambda_n(E_e, \vec{k}_e, \vec{k}_{\bar{\nu}}, \vec{\xi}_n,
  \vec{\xi}_e)}{dE_e d\Omega_e d\Omega_{\bar{\nu}}}\Big|_{(\rm
  NLO)}\nonumber\\
\hspace{-0.15in}&& + \sum^4_{m = 1}\frac{d^5
  \lambda^{(m)}_n(E_e, \vec{k}_e, \vec{k}_{\bar{\nu}}, \vec{\xi}_n,
  \vec{\xi}_e)}{dE_e d\Omega_e d\Omega_{\bar{\nu}}}\Big|_{(\rm N^2LO)},
\end{eqnarray}
where we have followed the notation \cite{Ivanov2013, Ivanov2017,
  Ivanov2018, Ivanov2019, Ivanov2020}.  Then, $\vec{\xi}_n$ and
$\vec{\xi}_e$ are unit vectors of spin--polarization of the neutron
and electron \cite{Ivanov2017, Ivanov2018} (see also
\cite{Itzykson1980}), respectively, $(E_e, \vec{k}_e)$ and
$(E_{\bar{\nu}}, \vec{k}_{\bar{\nu}})$ are the energies and 3--momenta
of the electron and antineutrino, respectively, $d\Omega_e$ and
$d\Omega_{\bar{\nu}}$ are infinitesimal solid angles in the directions
of electron $\vec{k}_e$ and antineutrino $\vec{k}_{\bar{\nu}}$
3--momenta, respectively, $E_0 = (m^2_n - m^2_p + m^2_e)/2m_n =
1.2926\,{\rm MeV}$ is the end--point energy of the electron--energy
spectrum \cite{Ivanov2013}, $F(E_e, Z = 1)$ is the relativistic Fermi
function equal to (see, for example, \cite{Wilkinson1982} and a
discussion in \cite{Ivanov2017})
\begin{eqnarray}\label{eq:3}
\hspace{-0.3in}F(E_e, Z = 1 ) = \Big(1 +
\frac{1}{2}\gamma\Big)\,\frac{4(2 r_pm_e\beta)^{2\gamma}}{\Gamma^2(3 +
  2\gamma)}\,\frac{\displaystyle e^{\,\pi \alpha/\beta}}{(1 -
  \beta^2)^{\gamma}}\,\Big|\Gamma\Big(1 + \gamma + i\,\frac{\alpha
}{\beta}\Big)\Big|^2,
\end{eqnarray}
where $\beta = k_e/E_e = \sqrt{E^2_e - m^2_e}/E_e$ is the electron
velocity, $\gamma = \sqrt{1 - \alpha^2} - 1$, $r_p$ is the electric
radius of the proton.  In the numerical calculations we use $r_p =
0.841\,{\rm fm}$ \cite{Pohl2010, Antognini2013}. The correlation
function $\zeta(E_e)$, responsible for the correct value of the
neutron lifetime, and the correlation coefficients $X(E_e)$, where $X
= {\rm electron-antineutrino\;correlation}\;a, {\rm electron\;
  asymmetry}\; A, {\rm antineutrino\; asymmetry}\; B$, and so on, have
the following structure
\begin{eqnarray}\label{eq:4}
\hspace{-0.3in}\zeta(E_e) &=& \zeta(E_e)_{(\rm LO)} + \zeta(E_e)_{(\rm
  NLO)} + \zeta(E_e)_{(\rm N^2LO)}, \nonumber\\             
\hspace{-0.3in}X(E_e) &=& X(E_e)_{(\rm LO)} + X(E_e)_{(\rm NLO)} +
X(E_e)_{(\rm N^2LO)},
\end{eqnarray}
where the abbreviations ${\rm LO}$, ${\rm NLO}$ and ${\rm N^2LO}$ mean
``leading-order'', ``next-to-leading-order'' and
``next-to-next-to-leading-order'' in the large nucleon mass $m_N$
expansion, respectively. With respect to an expansion in powers of
$1/m_N$ these terms are of order $O(1)$, $O(1/m_N)$ and $O(1/m^2_N)$,
respectively.

The correlation function $\zeta(E_e)$ and correlation coefficients
$X(E_e)_{(\rm LO)}$, calculated to leading order in the large nucleon
mass $m_N$ expansion, contain also the contributions of radiative
corrections of order $O(\alpha/\pi)$. A complete set of corrections of
order $10^{-3}$, defined by radiative corrections of order
$O(\alpha/\pi)$ and corrections of order $O(E_e/m_N)$, caused by weak
magnetism and proton recoil, which we denote as $\zeta(E_e)_{(\rm
  NLO)}$ and $X(E_e)_{(\rm NLO)}$ to the neutron lifetime and to the
correlation coefficients with correlation structure independent of the
electron spin $\vec{\zeta}_e$ were calculated in \cite{Sirlin1967,
  Shann1971, Bilenky1959, Wilkinson1982, Sirlin1986, Sirlin2004,
  Sirlin2006, Seng2018, Seng2018a, Sirlin2019} (see also
\cite{Ando2004, Gudkov2006, Ivanov2013}). In turn, a complete set of
corrections of order of  $10^{-3}$, including the radiative corrections of
order $O(\alpha/\pi)$ and the weak magnetism and proton recoil
corrections of order $O(E_e/m_N)$, to the correlation coefficients,
caused by correlations with the electron spin $\vec{\zeta}_e$, have
been calculated in \cite{Ivanov2017, Ivanov2018, Ivanov2019}. The
radiative corrections of order $O(\alpha E_e/m_N)$, which can be also
added to $\zeta(E_e)_{(\rm NLO)}$ and $X(E_e)_{(\rm NLO)}$, have been
recently calculated in \cite{Ivanov2019a, Ivanov2020a}. The
correlation coefficient $b(E_e)$ is the Fierz interference term
\cite{Fierz1937}. The structure and the value of the Fierz
interference term may depend on interactions beyond the SM. A recent
information of the theoretical and experimental status of the Fierz
interference term can be found in \cite{Abele2019, Sun2020, Hardy2020,
  Severijns2019, Ivanov2019b}.

The last two terms in Eq.(\ref{eq:2}) determine the contributions to
the electron--energy and angular distributions of order $O(E_e/m_N)$
and $O(E^2_e/m^2_N)$, respectively, induced by weak magnetism and
proton recoil, with the correlation structure, which cannot be reduced
to the correlation structures by Jackson {\it et al.}
\cite{Jackson1957}. We give the analytical expressions of these terms
in Appendix B.

Below we propose a detailed analysis of the structure and calculation
of the corrections $\zeta(E_e)_{(\rm N^2LO)}$ and $X(E_e)_{(\rm
  N^2LO)}$ to the neutron lifetime and the correlation coefficients of
the neutron beta decay, respectively. For the correction
$\zeta(E_e)_{(\rm N^2LO)}$ we define the following structure
\begin{eqnarray}\label{eq:5}
\hspace{-0.3in}\zeta(E_e)_{(\rm N^2LO)} = \zeta^{(1)}(E_e)_{(\rm
  N^2LO)} + \zeta^{(2)}(E_e)_{(\rm N^2LO)} + \zeta^{(3)}(E_e)_{(\rm
  N^2LO)} + \zeta^{(4)}(E_e)_{(\rm N^2LO)},
\end{eqnarray}
where i) $\zeta^{(1)}(E_e)_{(\rm N^2LO)}$ is induced by $\delta {\cal
  M}_n$ in the amplitude of the neutron beta decay (see
Eq.(\ref{eq:A.14}) and Eq.(\ref{eq:A.15})), ii)
$\zeta^{(2)}(E_e)_{(\rm N^2LO)}$ is caused by corrections of order
$O(E^2_e/m^2_N)$ to the phase--volume of the neutron beta decay (see
Eq.(\ref{eq:A.21})); iii) $\zeta^{(3)}(E_e)_{(\rm N^2LO)}$ is the
correction, obtained by the quadratic and crossing terms of order
$O(E_e/m_N)$ in the amplitude of the neutron beta decay
Eq.(\ref{eq:A.14}) without $\delta {\cal M}_n$, and iv)
$\zeta^{(4)}(E_e)_{(\rm N^2LO)}$ is determined by the corrections of
order $O(E_e/m_N)$ to the electron--energy and angular distribution of
the neutron beta decay with the account for the contributions of order
$O(E_e/m_N)$ from the phase--volume of the neutron beta decay (see
Eq.(\ref{eq:A.21})).

In order to define the structure of the corrections $X(E_e)_{(\rm
  N^2LO)}$ to the correlation coefficients $X(E_e) = a(E_e), A(E_e)$
and so on we remind that the correlation coefficients $X(E_e)$ appear
in the electron--energy and angular distribution of the neutron beta
decay in the form (see, for example, \cite{Ivanov2013})
\begin{eqnarray}\label{eq:6}
\hspace{-0.3in}\zeta(E_e)X(E_e) = \bar{X}(E_e)_{(\rm LO)} +
\bar{X}(E_e)_{(\rm NLO)} + \bar{X}(E_e)_{(\rm N^2LO)},
\end{eqnarray}
where we have added the term $\bar{X}(E_e)_{(\rm N^2LO)}$, which we
calculate in this paper, to well--known first two terms
$\bar{X}(E_e)_{(\rm LO)}$ and $\bar{X}(E_e)_{(\rm NLO)}$ (see
\cite{Abele2008, Nico2009, Bilenky1959, Gudkov2006, Ivanov2013,
  Ivanov2017, Ivanov2018, Ivanov2019}). All terms in the
right-hand-side (r.h.s.) of Eq.(\ref{eq:6}) are calculated from
$|{\cal M}_n|^2$ (see, for example, \cite{Ivanov2013, Ivanov2017,
  Ivanov2018, Ivanov2019}), where ${\cal M}_n$ is given in
Eq.(\ref{eq:A.14}), for a polarized neutron, a polarized electron and
an unpolarized proton by taking into account the contributions of the
phase--volume of the neutron beta decay (see Eq.(\ref{eq:A.21})) at
the neglect of the contributions proportional to $1/m^3_N$ and
$1/m^4_N$. In this case the corrections $\bar{X}(E_e)_{(\rm N^2LO)}$
have the structure of the correlation function $\zeta(E_e)_{(\rm
  N^2LO)}$ (see Eq.(\ref{eq:5}))
\begin{eqnarray}\label{eq:7}
\hspace{-0.3in}\bar{X}(E_e)_{(\rm N^2LO)} = \bar{X}^{(1)}(E_e)_{(\rm
  N^2LO)} + \bar{X}^{(2)}(E_e)_{(\rm N^2LO)} +
\bar{X}^{(3)}(E_e)_{(\rm N^2LO)} + \bar{X}^{(4)}(E_e)_{(\rm N^2LO)}.
\end{eqnarray}
At the neglect of the contributions of the radiative corrections of
order $O(\alpha/\pi)$ and $O(\alpha E_e/m_N)$ the correlation function
$\zeta(E_e)$ has the structure
\begin{eqnarray}\label{eq:8}
\hspace{-0.3in}\zeta(E_e) = 1 + \zeta(E_e)_{(\rm NLO)} +
\zeta(E_e)_{(\rm N^2LO)}
\end{eqnarray}
with respect to the large nucleon mass $m_N$ expansion, where we have
added the term $\zeta(E_e)_{(\rm N^2LO)}$ to the first two terms (see
Eq.(7) of Ref.\cite{Ivanov2013}). Taking into Eq.(\ref{eq:8}) we
define the correlation coefficients $X(E_e)$ as follows
\begin{eqnarray}\label{eq:9}
\hspace{-0.3in}X(E_e) = \frac{\bar{X}(E_e)_{(\rm LO)} +
  \bar{X}(E_e)_{(\rm NLO)} + \bar{X}(E_e)_{(\rm N^2LO)}}{ 1 +
  \zeta(E_e)_{(\rm NLO)} + \zeta(E_e)_{(\rm N^2LO)}}.
\end{eqnarray}
Expanding the r.h.s of Eq.(\ref{eq:9}) in powers of $1/m_N$ and
restricting such an expansion by the contributions proportional to
$1/m^2_N$ we get 
\begin{eqnarray}\label{eq:10}
  \hspace{-0.3in}X(E_e) = X(E_e)_{(\rm LO)} + X(E_e)_{(\rm NLO)} +
  X(E_e)_{(\rm N^2LO)},
\end{eqnarray}
where $X(E_e)_{(\rm LO)} = \bar{X}(E_e)_{(\rm LO)}$ and $X(E_e)_{(\rm
  NLO)} = \bar{X}(E_e)_{(\rm NLO)} - \bar{X}(E_e)_{(\rm
  LO)}\zeta(E_e)_{(\rm NLO)}$. The exact expressions for $X(E_e)_{(\rm
  LO)}$ and $X(E_e)_{(\rm NLO)}$ have been calculated in
\cite{Bilenky1959, Gudkov2006, Ivanov2013, Ivanov2017, Ivanov2018,
  Ivanov2019}. The correction $X(E_e)_{(\rm N^2LO)}$ has the structure
\begin{eqnarray}\label{eq:11}
\hspace{-0.3in}X(E_e)_{(\rm N^2LO)} = \bar{X}(E_e)_{(\rm N^2LO)} -
\bar{X}(E_e)_{(\rm LO)}\zeta(E_e)_{(\rm N^2LO)} + \bar{X}(E_e)_{(\rm
  LO)} \zeta^2(E_e)_{(\rm NLO)} - \bar{X}(E_e)_{(\rm
  NLO)}\zeta(E_e)_{(\rm NLO)}.
\end{eqnarray}
For the calculation of the last two terms we use the results for the
corrections $\zeta(E_e)_{(\rm NLO)}$ and $\bar{X}(E_e)_{(\rm NLO)}$,
obtained in \cite{Ivanov2013, Ivanov2017, Ivanov2018, Ivanov2019},
which we adduce in Appendix C for completeness of our analysis.

\section{Analytical expressions for complete set of corrections
  defining $X(E_e)_{(\rm N^2LO)}$ in Eq.(\ref{eq:11})}
\label{sec:barX}

Following \cite{Ivanov2013, Ivanov2017, Ivanov2018, Ivanov2019,
  Ivanov2020} we calculate the complete set of corrections defining
the total corrections $\zeta(E_e)_{(\rm N^2LO)}$ and $X(E_e)_{(\rm
  N^2LO)}$ to the correlation function $\zeta(E_e)$ and correlation
coefficients $X(E_e)$ for $X(E_e) = a(E_e), A(E_e)$ and so on.  We
would like to notice that all corrections of order $O(E^2_e/m^2_N)$ to
the correlation coefficients $D(E_e)$, $R(E_e)$ and $L(E_e)$ are
either equal to zero or much smaller than $10^{-5}$.  The corrections
to the Fierz interference term $b(E_e)$ we do not consider, since up
to now its experimental status is not well defined \cite{Abele2019,
  Sun2020, Hardy2020, Severijns2019, Ivanov2019b}.

\subsection*{\bf 1. Corrections  $\zeta^{(1)}(E_e)_{(\rm N^2LO)}$ and
$\bar{X}^{(1)}(E_e)_{(\rm N^2LO)}$, induced by the terms of order
$O(E^2_0/m^2_N)$ in the amplitude of the neutron beta decay
Eq.(\ref{eq:A.14}) and Eq.(\ref{eq:A.15})}

Using the standard procedure for the calculation of the
electron--energy and angular distribution of the neutron beta decay
(see, for example, Appendix A in \cite{Ivanov2013}) we obtain the
following analytical expressions for the corrections
$\zeta^{(1)}(E_e)_{(\rm N^2LO)}$ and $\bar{X}^{(1)}(E_e)_{(\rm
  N^2LO)}$:
\begin{eqnarray*}
\hspace{-0.21in}&& \zeta^{(1)}(E_e)_{(\rm N^2LO)} = \frac{1}{1 + 3 g^2_A}
\frac{E^2_0}{4 m^2_N} \Big\{\Big[\big(- 6 g^2_A + 4 \kappa + 2\big) -
  \frac{2}{3}\,\big(g_A + 2 \kappa + 1\big) \Big(1 -
  \frac{m^2_e}{E^2_e}\Big)\Big] \frac{E_e}{E_0}\Big(1 -
\frac{E_e}{E_0}\Big) \nonumber\\
\hspace{-0.21in}&&   \bar{a}^{(1)}(E_e)_{(\rm N^2LO)} = \frac{1}{1 + 3 g^2_A}
\frac{E^2_0}{4 m^2_N} \Big\{ 8 g^2_A \frac{E_e}{E_0}\Big(1 -
\frac{E_e}{E_0}\Big) + ( - g^2_A + 2 \kappa + 1)\,
\frac{m^2_e}{E^2_0}\Big\}, \nonumber\\
\hspace{-0.21in}&&\bar{A}^{(1)}(E_e)_{(\rm N^2LO)} = \frac{1}{1 + 3 g^2_A}
\frac{E^2_0}{4 m^2_N} \Big\{ - 4 g_A (g_A +
\kappa)\,\frac{E_e}{E_0}\Big(1 - \frac{E_e}{E_0}\Big) + 2 g_A( - g_A +
\kappa + 1)\, \frac{m^2_e}{E^2_0}\Big\}, \nonumber\\
\end{eqnarray*}
\begin{eqnarray}\label{eq:12}
\hspace{-0.21in}&& \bar{B}^{(1)}(E_e)_{(\rm N^2LO)} = \frac{1}{1 + 3
  g^2_A} \frac{E^2_0}{4 m^2_N}\Big\{ - 4 g_A (g_A -
\kappa)\,\frac{E_e}{E_0}\Big(1 - \frac{E_e}{E_0}\Big) - 2 \kappa\,
\frac{m_e}{E_0}\frac{m_e}{E_e} + 2 g_A(g_A + \kappa + 1)\,
\frac{m^2_e}{E^2_0}\Big\}, \nonumber\\
\hspace{-0.21in}&& \bar{K}^{(1)}_n(E_e)_{(\rm N^2LO)} = \frac{1}{1 + 3
  g^2_A} \frac{E^2_0}{4 m^2_N} \Big\{(- 4 g_A(g_A + \kappa)\,
\frac{E_e}{E_0}\Big(1 - \frac{E_e}{E_0}\Big)\Big\},\nonumber\\
\hspace{-0.21in}&&  \bar{Q}^{(1)}_n(E_e)_{(\rm N^2LO)} = \frac{1}{1 + 3
  g^2_A} \frac{E^2_0}{4 m^2_N} \Big\{4 g_A (g_A - \kappa)\,
\frac{E_e}{E_0}\Big(1 - \frac{E_e}{E_0}\Big)\Big\}, \nonumber\\
\hspace{-0.21in}&&   \bar{G}^{(1)}(E_e)_{(\rm N^2LO)} = \frac{1}{1 + 3 g^2_A}
\frac{E^2_0}{4 m^2_N} \Big\{2 (3 g^2_A - 2 \kappa -
1)\,\frac{E_e}{E_0}\Big(1 - \frac{E_e}{E_0}\Big) + \frac{2}{3} g^2_A
\frac{E_e}{E_0}\Big(1 - \frac{E_e}{E_0}\Big) \Big(1 -
\frac{m_e}{E_0}\Big) \nonumber\\
\hspace{-0.21in}&& + \frac{2}{3}( g^2_A + 2 \kappa + 1)\,
\frac{m_e}{E_0}\Big(1 - \frac{E_e}{E_0}\Big)\Big\}, \nonumber\\
\hspace{-0.21in}&& \bar{H}^{(1)}(E_e)_{(\rm N^2LO)} = \frac{1}{1 + 3
  g^2_A} \frac{E^2_0}{4 m^2_N}\, \frac{m_e}{E_e} \Big\{ - 2 (g^2_A + 2
\kappa + 1)\, \frac{E_e}{E_0}\Big(1 - \frac{E_e}{E_0}\Big)+ 2 \kappa\,
\frac{E_e}{E_0} + (g^2_A - 2\kappa - 1)\, \frac{m^2_e}{E^2_0} \Big\},
\nonumber\\
\hspace{-0.21in}&& \bar{N}^{(1)}(E_e)_{(\rm N^2LO)} = \frac{1}{1 + 3
  g^2_A} \frac{E^2_0}{4 m^2_N} \, \frac{m_e}{E_e} \Big\{- 4 g_A (g_A +
\kappa)\,\frac{E_e}{E_0}\Big(1 - \frac{E_e}{E_0}\Big) + 2 \kappa g_A
\frac{E_e}{E_0} + 2 g_A(g_A - \kappa - 1)\, \frac{m^2_e}{E^2_0}
\Big\}, \nonumber\\
\hspace{-0.21in}&& Q^{(1)}_e(E_e)_{(\rm N^2LO)} = \frac{1}{1 + 3
   g^2_A} \frac{E^2_0}{4 m^2_N} \Big\{- 4 g_A (g_A + \kappa)\,
 \frac{E_e}{E_0}\Big(1 - \frac{E_e}{E_0}\Big) + \frac{2}{3} g_A (2 g_A
 + 4 \kappa + 1)\, \frac{E_e}{E_0} \Big(1 - \frac{E_e}{E_0}\Big)
 \Big(1 + \frac{m_e}{E_e}\Big) \nonumber\\
\hspace{-0.21in}&&+ 2 \kappa g_A\, \frac{E_e}{E_0} + 2 g_A (g_A -
\kappa - 1)\, \frac{m^2_e}{E^2_0} \Big\}, \nonumber\\
\hspace{-0.21in}&& \bar{K}^{(1)}_e(E_e)_{(\rm N^2LO)} = \frac{1}{1 +
   3 g^2_A} \frac{E^2_0}{4 m^2_N} \Big\{ - 2 (g^2_A + 2 \kappa + 1)\,
 \frac{E_e}{E_0}\Big(1 - \frac{E_e}{E_0}\Big) + 2(- 3 g^2_A + 2 \kappa
 + 1)\, \frac{E_e}{E_0}\Big(1 - \frac{E_e}{E_0}\Big)\nonumber\\
 \hspace{-0.21in}&&\times \Big(1 + \frac{m_e}{E_e}\Big) + \Big( - 2
 \kappa + (g^2_A - 2 \kappa - 1)\, \frac{m_e}{E_0}\Big)\,
 \frac{m_e}{E_0}\Big\}.
\end{eqnarray}
These corrections are proportional to the factor $(E^2_0/4 m^2_N)/(1 +
3 g^2_A) = 8.05\times 10^{-8} \sim 10^{-7}$, calculated for $E_0 =
1.2926\,{\rm MeV}$, $m_N = (m_n + m_p)/2 = 938.9188$ and $g_A =
1.2764$ \cite{Abele2018}. Because of such a factor this sort of
corrections are of oder $10^{-6}$ and smaller. They can be neglected
in comparison with corrections of order of  $10^{-5}$.

\subsection*{\bf 2. Corrections $\zeta^{(2)}(E_e)_{(\rm N^2LO)}$ and
  $\bar{X}^{(2)}(E_e)_{(\rm N^2LO)}$, induced by
  next-to-next-to-leading order terms from phase--volume of the
  neutron beta decay (see Eq.(\ref{eq:A.21})}

The corrections $\zeta^{(2)}(E_e)_{(\rm N^2LO)}$ and
$\bar{X}^{(2)}(E_e)_{(\rm N^2LO)}$, induced by next-to-next-to-leading
order terms in the large nucleon mass $m_N$ expansion of the
phase--volume of the neutron beta decay (see Eq.(\ref{eq:A.21}), are
given by
\begin{eqnarray*}
\hspace{-0.21in}&& \zeta^{(2)}(E_e)_{(\rm N^2LO)} = 6\,
\frac{E^2_e}{m^2_N}\Big\{ \Big(1 - \frac{1}{4}\,\frac{E_0}{E_e}\Big) +
\frac{1}{3} \Big[1- 2\, \frac{1 - g^2_A}{1 + 3 g^2_A}\, \Big( 1 -
  \frac{1}{8}\,\frac{E_0}{E_e}\Big)\Big] \Big(1 -
\frac{m^2_e}{E^2_e}\Big) \Big\}, \nonumber\\
\hspace{-0.21in}&& \bar{a}^{(2)}(E_e)_{(\rm N^2LO)} = 12\,
\frac{E^2_e}{m^2_N}\Big\{ - \Big(1 - \frac{1}{8}\,
\frac{E_0}{E_e}\Big) + \frac{1}{3}\, \frac{1 - g^2_A}{1 + 3 g^2_A}\,
\Big(1 - \frac{1}{4}\, \frac{E_0}{E_e}\Big)\Big\}, \nonumber\\
\hspace{-0.21in}&& \bar{A}^{(2)}(E_e)_{(\rm N^2LO)} = 12 \,
\frac{E^2_e}{m^2_N}\, \Big\{ \frac{g_A(1 - g_A)}{1 + 3 g^2_A}\, \Big(1
- \frac{1}{4}\, \frac{E_0}{E_e}\Big) + \frac{1}{3}\,\frac{g_A(1 -
  g_A)}{1 + 3 g^2_A}\Big(1 -
\frac{m^2_e}{E^2_e}\Big)\Big\},\nonumber\\
\hspace{-0.21in}&&   \bar{B}^{(2)}(E_e)_{(\rm N^2LO)} = 12 \,
\frac{E^2_e}{m^2_N}\, \Big\{ \frac{g_A(1 + g_A)}{1 + 3 g^2_A}\, \Big(1
- \frac{1}{4}\, \frac{E_0}{E_e}\Big)\Big\}, \nonumber\\
\end{eqnarray*}
\begin{eqnarray}\label{eq:13}
\hspace{-0.21in}&& \bar{K}^{(2)}_n(E_e)_{(\rm N^2LO)} = 24\,
\frac{E^2_e}{m^2_N}\,\Big\{ - \frac{g_A(1 - g_A)}{1 + 3 g^2_A}\,
\Big(1 - \frac{1}{8}\, \frac{E_0}{E_e}\Big)\Big\}, \nonumber\\
\hspace{-0.21in}&& \bar{Q}^{(2)}_n(E_e)_{(\rm N^2LO)} = 24\,
\frac{E^2_e}{m^2_N}\,\Big\{ - \frac{g_A(1 + g_A)}{1 + 3 g^2_A}\,
\Big(1 - \frac{1}{8}\, \frac{E_0}{E_e}\Big)\Big\}, \nonumber\\
\hspace{-0.21in}&& \bar{G}^{(2)}(E_e)_{(\rm N^2LO)} = 6\,
\frac{E^2_e}{m^2_N}\, \Big\{- \Big(1 -
\frac{1}{4}\,\frac{E_0}{E_e}\Big) - \frac{1}{3}\, \Big(1 -
\frac{m^2_e}{E^2_e}\Big) + \frac{2}{3}\, \frac{1 - g^2_A}{1 + 3
  g^2_A}\, \Big(1 - \frac{1}{8}\, \frac{E_0}{E_e}\Big) \Big\},
\nonumber\\
\hspace{-0.21in}&& \bar{H}^{(2)}(E_e)_{(\rm N^2LO)} = 6\,
\frac{E^2_e}{m^2_N}\,\Big\{ - \frac{1 - g^2_A}{1 + 3 g^2_A}\,
\frac{m_e}{E_e}\, \Big(1 - \frac{1}{4}\,\frac{E_0}{E_e}\Big)
\Big\},\nonumber\\
\hspace{-0.21in}&& \bar{N}^{(2)}(E_e)_{(\rm N^2LO)} = 12\,
\frac{E^2_e}{m^2_N}\, \frac{m_e}{E_e}\,\Big\{ - \frac{g_A(1 - g_A)}{1
  + 3 g^2_A}\, \Big(1 - \frac{1}{4}\, \frac{E_0}{E_e}\Big) -
\frac{1}{3}\,\frac{g_A(1 - g_A)}{1 + 3 g^2_A}\, \Big(1 -
\frac{m^2_e}{E^2_e}\Big)\Big\}, \nonumber\\
 \hspace{-0.21in}&& \bar{Q}^{(2)}_e(E_e)_{(\rm N^2LO)} = 12\,
 \frac{E^2_e}{m^2_N}\, \Big\{ - \frac{g_A (1 - g_A)}{1 + 3 g^2_A}\,
 \Big(1 - \frac{1}{4}\, \frac{E_0}{E_e}\Big) - \frac{1}{3}\,\frac{g_A
   (1 - g_A)}{1 + 3 g^2_A}\,\Big(1 - \frac{m^2_e}{E^2_e}\Big) +
 \frac{2}{3}\,\frac{g_A (1 + g_A)}{1 + 3 g^2_A}\nonumber\\
 \hspace{-0.21in}&& \times \, \Big(1 - \frac{1}{8}\,
 \frac{E_0}{E_e}\Big)\Big(1 + \frac{m_e}{E_e}\Big)\Big\}, \nonumber\\
\hspace{-0.21in}&& \bar{K}^{(2)}_e(E_e)_{(\rm N^2LO)} = 12\,
 \frac{E^2_e}{m^2_N}\,\Big\{ \Big(1 - \frac{1}{8}\,
 \frac{E_0}{E_e}\Big)\Big(1 + \frac{m_e}{E_e}\Big) - \frac{1}{2}\,
 \frac{1 - g^2_A}{1 + 3 g^2_A}\, \Big(1 - \frac{1}{4}\,
 \frac{E_0}{E_e}\Big)\Big\}.
\end{eqnarray}
These corrections are proportional to the factors $6 E^2_0/m^2_N =
1.14 \times 10^{-5}$, $12 E^2_0/m^2_N = 2.28 \times 10^{-5}$ and $24
E^2_0/m^2_N = 4.56 \times 10^{-5}$. This means that the corrections
$O(E^2_e/m^2_N)$, induced by the phase--volume of the neutron beta
decay, such as $\zeta^{(2)}(E_e)_{(\rm N^2LO)}$ and
$\bar{X}^{(2)}(E_e)_{(\rm N^2LO)}$ for $X(E_e) = a(E_e), B(E_e),
Q_n(E_e), G(E_e)$ and $K_e(E_e)$ are of the required order
$10^{-5}$. We define these corrections as polynomials in the variable
$E_e/E_0$ with the numerical analysis of the coefficients
\begin{eqnarray}\label{eq:14}
\hspace{-0.3in}\zeta^{(2)}(E_e)_{(\rm N^2LO)} &=& 1.60 \times
10^{-5}\, \frac{E^2_e}{E^2_0}\;,\; \bar{a}^{(2)}(E_e)_{(\rm N^2LO)} =
- 2.36 \times 10^{-5}\,\frac{E^2_e}{E^2_0}\;,\;
\bar{B}^{(2)}(E_e)_{(\rm N^2LO)} = 1.12 \times 10^{-5}\,
\frac{E^2_e}{E^2_0},\nonumber\\
\hspace{-0.3in}\bar{Q}^{(2)}_n(E_e)_{(\rm N^2LO)} &=&
- 2.25 \times 10^{-5}\,
\frac{E^2_e}{E^2_0}\;,\; \bar{G}^{(2)}(E_e)_{(\rm N^2LO)} =
- 1.60 \times
10^{-5}\,\frac{E^2_e}{E^2_0}\;,\;\bar{K}^{(2)}_e(E_e)_{(\rm N^2LO)} =
2.40 \times 10^{-5}\, \frac{E^2_e}{E^2_0},\nonumber\\
\hspace{-0.3in}&&
\end{eqnarray}
where we have neglected the contributions of the terms of a few part
of $10^{-6}$. We would like to notice that, for example, the
corrections $\bar{B}^{(2)}(E_e)$ and $\bar{Q}^{(2)}_n(E_e)$ should be
important for experimental searches of interactions beyond the SM by
measuring antineutrino and electron asymmetries of the neutron beta
decay with experimental sensitivities at the level of a few parts of
$10^{5}$ or even better \cite{Abele2016, Abele2018}.

The other corrections in Eq.(\ref{eq:13}) such as
$\bar{A}^{(2)}(E_e)_{(\rm N^2LO)}$ and so on are proportional to the
factors $g_A(1 - g_A)/(1 + 3 g^2_A) = - 0.060$ and $(1 - g^2_A)/(1 + 3
g^2_A) = - 0.107$ for $g_A = 1.2764$ \cite{Abele2018}, which reduce
the values of these corrections to the level of a few parts of
$10^{-6}$ or even smaller. We neglect such contributions with respect
to contributions of order of  $10^{-5}$ here and below in all
next-to-next-to-leading order corrections in the large nucleon mass
$m_N$ expansion.

\subsection*{\bf 3. Corrections $\zeta^{(3)}(E_e)_{(\rm N^2LO)}$ and
  $\bar{X}^{(3)}(E_e)_{(\rm N^2LO)}$, induced by quadratic and
  crossing terms of order $O(E_e/m_N)$ in the amplitude of the neutron
  beta decay Eq.(\ref{eq:A.14}) without $\delta {\cal M}_n$}

The analytical expressions of corrections $O(E^2_e/m^2_N)$, induced by
quadratic and crossing terms of order $O(E_e/m_N)$ to the amplitude of
the neutron beta decay Eq.(\ref{eq:A.14}) without $\delta {\cal M}_n$,
are equal to
\begin{eqnarray*}
\hspace{-0.21in}&& \zeta^{(3)}(E_e)_{(\rm N^2LO)} = \frac{1}{1 + 3
  g^2_A}\, \frac{E^2_0}{4 m^2_N} \Big\{ 3 g^2_A\Big(1 +
\frac{1}{3}\,\frac{m^2_e}{E^2_0}\Big) - 2 g_A\big(g_A - 2(\kappa +
1)\big)\,\Big(1 - \frac{E_e}{E_0}\Big) - 2 g_A\big(g_A + 2(\kappa +
1)\big)\nonumber\\
\hspace{-0.21in}&& \times \, \frac{E_e}{E_0}\Big(1 -
\frac{m^2_e}{E^2_e}\Big) + \big(g^2_A + 2 (\kappa + 1)^2\big)\, \Big(1
- \frac{E_e}{E_0}\Big)^2 - \big(g^2_A + 2 (\kappa +
1)^2\big)\,\frac{E^2_e}{E^2_0} \Big(1 - \frac{m^2_e}{E^2_e}\Big) -
\frac{2}{3}\,\big(g^2_A + 4 (\kappa + 1)^2\big)\nonumber\\
\hspace{-0.21in}&& \times \, \frac{E_e}{E_0}\Big(1 -
\frac{E_e}{E_0}\Big) \Big(1 - \frac{m^2_e}{E^2_e}\Big)\Big\},
\nonumber\\
\hspace{-0.21in}&& \bar{a}^{(3)}(E_e)_{(\rm N^2LO)} = \frac{1}{1 + 3
  g^2_A}\, \frac{E^2_0}{4 m^2_N}\Big\{- g^2_A \Big(1 - 2 (\kappa +
1)^2 \frac{m^2_e}{E^2_0}\Big) - \big(g^2_A + 2 (\kappa + 1)^2\big)\,
\frac{E^2_e}{E^2_0} - 2 g_A \big(g_A - 2 (\kappa + 1)\big)\,
\frac{E_e}{E_0} \nonumber\\
  \end{eqnarray*}
\begin{eqnarray}\label{eq:15} 
\hspace{-0.21in}&& - 2 g_A \big(g_A + 2 (\kappa + 1)\big)\Big(1 -
\frac{E_e}{E_0}\Big) + 2 \big(g^2_A + 2 (\kappa +
1)^2\big)\,\frac{E_e}{E_0} \Big(1 - \frac{E_e}{E_0}\Big) - \big(g^2_A
+ 2 (\kappa + 1)^2\big)\,\Big(1 - \frac{E_e}{E_0}\Big)^2 \Big\},
\nonumber\\
  \hspace{-0.21in}&& \bar{A}^{(3)}(E_e)_{(\rm N^2LO)} = \frac{1}{1 + 3
    g^2_A} \frac{E^2_0}{4 m^2_N} \Big\{- 2 \Big(g^2_A - \big(g^2_A -
  (\kappa + 1)^2\big)\, \frac{m^2_e}{E^2_0}\Big) + 4 g_A (\kappa + 1)\,
  \frac{E_e}{E_0} - 2 (\kappa + 1)^2\, \frac{E^2_e}{E^2_0} \nonumber\\
\hspace{-0.21in}&& + 2 g_A \big(g_A - (\kappa + 1)\big) \, \Big(1 -
\frac{E_e}{E_0}\Big) - 2 (\kappa + 1)\, \big(g_A - (\kappa + 1)\big)\,
\frac{E_e}{E_0} \Big(1 - \frac{E_e}{E_0}\Big) + 2 g_A (\kappa + 1)
\Big(1 - \frac{E_e}{E_0}\Big)^2 \Big\}, \nonumber\\
  \hspace{-0.21in}&& \bar{B}^{(3)}(E_e)_{(\rm N^2LO)} = \frac{1}{1 + 3
    g^2_A} \frac{E^2_0}{4 m^2_N} \Big\{2 g_A \Big(g_A - (\kappa + 1)\,
  \frac{m^2_e}{E^2_0}\Big) + 4 g_A (\kappa + 1)\, \frac{E_e}{E_0} - 2
  \kappa g_A \,\frac{E^2_e}{E^2_0} + 2 (\kappa + 1)^2 \nonumber\\  
  \hspace{-0.21in}&& \times \Big(1 - \frac{E_e}{E_0}\Big)^2 - 2 g_A
  \big(g_A + (\kappa + 2)\big)\, \frac{E_e}{E_0} \Big(1 -
  \frac{m^2_e}{E^2_e}\Big) - 2 g_A \big( \kappa g_A + (\kappa +
  1)^2\big)\, \frac{E_e}{E_0} \Big(1 - \frac{E_e}{E_0}\Big) \Big(1 -
  \frac{m^2_e}{E^2_e}\Big) \Big\}, \nonumber\\
\hspace{-0.21in}&& \bar{K}^{(3)}_n(E_e)_{(\rm N^2LO)} = \frac{1}{1 + 3
  g^2_A} \frac{E^2_0}{4 m^2_N}\, \Big\{2 \big(g^2_A - (\kappa + 1)^2\big)\,
\frac{E_e}{E_0} - 4 (\kappa + 1)\,\big(g_A - (\kappa + 1)\big)\,
\frac{E^2_e}{E^2_0}\Big\}, \nonumber\\
\hspace{-0.21in}&& \bar{Q}^{(3)}_n(E_e)_{(\rm N^2LO)} = \frac{1}{1 + 3
  g^2_A} \frac{E^2_0}{4 m^2_N} \Big\{- 2 g_A \big(g_A + (\kappa +
1)\big)\, \Big(1 - \frac{E_e}{E_0}\Big) + 2 (\kappa + 1)\big(g_A +
(\kappa + 1)\big)\,\frac{E_e}{E_0}\Big(1 - \frac{E_e}{E_0}\Big)
\nonumber\\
\hspace{-0.21in}&& - 2 (\kappa + 1)\big(g_A + (\kappa + 1)\big)\,
\Big(1 - \frac{E_e}{E_0}\Big)^2\Big\}, \nonumber\\
\hspace{-0.21in}&& \bar{G}^{(3)}(E_e)_{(\rm N^2LO)} = \frac{1}{1 + 3
  g^2_A}\, \frac{E^2_0}{4 m^2_N}\Big\{\Big(- 3 g^2_A + 2 \big(g^2_A +
(\kappa + 1)^2\big)\, \frac{m^2_e}{E^2_0}\Big) + 2 g_A \big(g_A + 2
(\kappa + 1)\big)\, \frac{E_e}{E_0} \nonumber\\
\hspace{-0.21in}&& - \big(g^2_A + 2 (\kappa + 1)^2\big)\,
\frac{E^2_e}{E^2_0} + 2 g_A \big(g_A - 2 (\kappa + 1)\big)\, \Big(1 -
\frac{E_e}{E_0}\Big) + \frac{2}{3}\, \big(g^2_A + (\kappa +
1)^2\big)\,\frac{E_e}{E_0}\Big(1 - \frac{E_e}{E_0}\Big) \nonumber\\
\hspace{-0.21in}&& - \big(g^2_A + 2 (\kappa + 1)^2\big)\,\Big(1 -
\frac{E_e}{E_0}\Big)^2 \Big\}, \nonumber\\
\hspace{-0.21in}&& \bar{H}^{(3)}(E_e)_{(\rm N^2LO)} = \frac{1}{1 + 3
  g^2_A}\, \frac{E^2_0}{4 m^2_N} \, \frac{m_e}{E_e}\Big\{ g^2_A \Big(1
- 2 \frac{m^2_e}{E^2_0} + \frac{E^2_e}{E^2_0}\Big) + 2 g_A \big(g_A +
2(\kappa + 1)\big)\, \Big(1 - \frac{E_e}{E_0}\Big)  \nonumber\\
\hspace{-0.21in}&& + \big(g^2_A + 2 (\kappa + 1)^2\big) \, \Big(1 -
\frac{E_e}{E_0}\Big)^2 \Big\}, \nonumber\\
\hspace{-0.21in}&& \bar{N}^{(3)}(E_e)_{(\rm N^2LO)} = \frac{1}{1 + 3
  g^2_A}\, \frac{E^2_0}{4 m^2_N}\, \frac{m_e}{E_e} \Big\{ 2 \Big(g^2_A
+ (\kappa + 1)(g_A - 1)\, \frac{m^2_e}{E^2_0}\Big) - 2 g_A
\,\frac{E_e}{E_0} - 2 (\kappa + 1)(g_A - 1)\, \frac{E^2_e}{E^2_0}
\nonumber\\
\hspace{-0.21in}&& - \frac{4}{3}\, g_A \big(g_A - 2 (\kappa +
1)\big)\, \Big(1 - \frac{E_e}{E_0}\Big) + \frac{2}{3}\, \big(g_A - 2
(\kappa + 1)\big)\, \frac{E_e}{E_0} \Big(1 - \frac{E_e}{E_0}\Big) -
\frac{2}{3}\, (\kappa + 1) \big(2 g_A - (\kappa + 1)\big)\, \Big(1 -
\frac{E_e}{E_0}\Big)^2 \Big\}, \nonumber\\
 \hspace{-0.21in}&& \bar{Q}^{(3)}_e(E_e)_{(\rm N^2LO)} = \frac{1}{1 +
   3 g^2_A}\, \frac{E^2_0}{4 m^2_N} \Big\{2 \Big(g^2_A - g_A (2 \kappa
 + 1)\, \frac{m_e}{E_0} + (2 g_A - 1)(\kappa + 1)\,
 \frac{m^2_e}{E^2_0}\Big) - 4 g_A (\kappa + 1)\, \frac{E_e}{E_0}
 \nonumber\\
\hspace{-0.21in}&&- 2 (2 g_A - 1) (\kappa + 1)\, \frac{E^2_e}{E^2_0} -
\frac{4}{3}\, g_A \big(g_A - 2 (\kappa + 1)\big)\, \Big(1 -
\frac{E_e}{E_0}\Big) - \frac{2}{3}\, (\kappa + 1)\big(2 g_A - (\kappa
+ 1)\big)\, \Big(1 - \frac{E_e}{E_0}\Big)^2 \nonumber\\
 \hspace{-0.21in}&& + \frac{2}{3}\,\big(g_A - 2 (\kappa + 1)\big)\,
 \frac{E_e}{E_0}\Big(1 - \frac{E_e}{E_0}\Big) + \frac{2}{3}\,(2 \kappa
 + 1)\big(g_A - 2 (\kappa + 1)\big)\, \frac{E_e}{E_0}\Big(1 -
 \frac{E_e}{E_0}\Big)\Big(1 + \frac{m_e}{E_e}\Big) + 2 (\kappa + 1)
 \nonumber\\
 \hspace{-0.21in}&& \times (g_A + \kappa)\, \frac{E^2_e}{E^2_0}\Big(1
 + \frac{m_e}{E_e}\Big)\Big\}, \nonumber\\
 \hspace{-0.21in}&& \bar{K}^{(3)}_e(E_e)_{(\rm N^2LO)} = \frac{1}{1 +
   3 g^2_A}\, \frac{E^2_0}{4 m^2_N}\,\Big\{g_A \Big(g_A + 2 \big(g_A -
 2 (\kappa + 1)\big)\, \frac{m_e}{E_0} - 2 g_A\,
 \frac{m^2_e}{E^2_0}\Big) + 2 g_A \big(g_A - 2 (\kappa + 1)\big)\,
 \frac{E_e}{E_0} \nonumber\\
 \hspace{-0.21in}&& + \big(g^2_A + 2 (\kappa + 1)^2\big)\,
 \frac{E^2_e}{E^2_0} + 2 (\kappa + 1)^2\,\frac{E^2_e}{E^2_0} \Big(1 +
 \frac{m_e}{E_e}\Big) - 2 \big(g^2_A + 2 (\kappa + 1)^2\big)\,
 \frac{E_e}{E_0} \Big(1 - \frac{E_e}{E_0}\Big) \Big(1 +
 \frac{m_e}{E_e}\Big) \nonumber\\
 \hspace{-0.21in}&& + 2 g_A \big(g_A + 2 (\kappa + 1)\big) \Big(1 -
 \frac{E_e}{E_0}\Big) + \big(g^2_A + 2 (\kappa + 1)^2\big)\, \Big(1 -
 \frac{E_e}{E_0}\Big)^2\Big\}.
\end{eqnarray}
These corrections are proportional to the factor $(E^2_0/4 m^2_N)/(1 +
3 g^2_A) = 8.05\times 10^{-8} \sim 10^{-7}$. In spite of a strong
dependence of on the axial coupling constant $g_A = 1.2764$ and the
isovector anomalous magnetic moment of the nucleon $\kappa = \kappa_p
- \kappa_n = 3.7059$ \cite{PDG2020} the value of the corrections in
Eq.(\ref{eq:15}) are at the level of a few parts of $10^{-6}$. We
neglect these contributions in our analysis of the corrections of order 
$O(E^2_e/m^2_N) \sim 10^{-5}$.

\subsection*{\bf 4. Corrections $\zeta^{(4)}(E_e)_{(\rm N^2LO)}$ and
  $\bar{X}^{(4)}(E_e)_{(\rm N^2L0)}$, obtained by multiplication of
  the corrections of order $O(E_e/m_N)$ to the electron--energy and
  angular distribution of the neutron beta decay by the corrections of
  order $O(E_e/m_N)$ from the phase--volume of the neutron beta decay
  (see Eq.(\ref{eq:A.21})}

Using the next-to-leading order corrections $\zeta(E_e)_{(\rm NLO)}$
and $\bar{X}(E_e)_{(\rm NLO)}$ in the large nucleon mass $m_N$
expansion $O(E_e/m_N)$, calculated in \cite{Ivanov2013, Ivanov2017,
  Ivanov2018, Ivanov2019} and adduced in Appendix C, we may calculate
the corrections $\zeta^{(4)}(E_e)_{(\rm N^2L0)}$ and
$\bar{X}^{(4)}(E_e)_{(\rm N^2L0)}$ multiplying the corrections
$\zeta(E_e)_{(\rm NLO)}$ and $\bar{X}(E_e)_{(\rm NLO)}$ by the terms
of order $O(E_e/m_N)$ from the phase--volume of the neutron beta decay
(see Eq.(\ref{eq:A.21})). We get
\begin{eqnarray}\label{eq:16}
\hspace{-0.3in}&& \zeta^{(4)}(E_e)_{(\rm N^2LO)} = 3\,
\frac{E_e}{m_N}\, \zeta(E_e)_{(\rm NLO)} - \frac{E_e}{m_N}\,
\bar{a}(E_e)_{(\rm NLO)} \Big(1 - \frac{m^2_e}{E^2_e}\Big) = -
1.97\times 10^{-5}\,\frac{E_e}{E_0} + 5.49 \times 10^{-5}\,
\frac{E^2_e}{E^2_0},\nonumber\\
\hspace{-0.3in}&& \bar{a}^{(4)}(E_e)_{(\rm N^2LO)} = 3\,
\frac{E_e}{m_N}\, \bar{a}(E_e)_{(\rm NLO)} - \frac{E_e}{m_N}\,
\zeta(E_e)_{(\rm NLO)} = 1.97 \times 10^{-5}\, \frac{E_e}{E_0} - 5.57
\times 10^{-5}\,\frac{E^2_e}{E^2_0},\nonumber\\
\hspace{-0.3in}&&\bar{A}^{(4)}(E_e)_{(\rm N^2LO)} = 3\,
\frac{E_e}{m_N}\, \bar{A}(E_e)_{(\rm NLO)} -
\frac{E_e}{m_N}\,\bar{K}_n(E_e)_{(\rm NLO)} \Big(1 -
\frac{m^2_e}{E^2_e}\Big) = - 1.31 \times
10^{-5}\,\frac{E^2_e}{E^2_0},\nonumber\\
  \hspace{-0.3in}&& \bar{B}^{(4)}(E_e)_{(\rm N^2LO)} = 3\,
  \frac{E_e}{m_N}\, \bar{B}(E_e)_{(\rm NLO)} = 
  - 1.47 \times 10^{-5}
  \,\frac{E_e}{E_0} + 3.91 \times 10^{-5}\,\frac{E^2_e}{E^2_0},
  \nonumber\\
\hspace{-0.3in}&& \bar{K}^{(4)}_n(E_e)_{(\rm N^2LO)} = 3\,
\frac{E_e}{m_N}\, \bar{K}_n(E_e)_{(\rm NLO)} - 3\,\frac{E_e}{m_N}\,
\bar{A}(E_e)_{(\rm NLO)} =
1.51 \times 10^{-5}\,\frac{E^2_e}{E^2_0}, \nonumber\\
\hspace{-0.3in}&& \bar{Q}^{(4)}_n(E_e)_{(\rm N^2LO)} = 3\,
\frac{E_e}{m_N}\, \bar{Q}_n(E_e)_{(\rm NLO)} - 3\, \frac{E_e}{m_N}\,
\bar{B}(E_e)_{(\rm NLO)} = 2.79 \times 10^{-5}\,\frac{E_e}{E_0} - 6.91
\times 10^{-5}\, \frac{E^2_e}{E^2_0},\nonumber\\
\hspace{-0.3in}&& \bar{G}^{(4)}(E_e)_{(\rm N^2LO)} = 3\,
\frac{E_e}{m_N}\, \bar{G}(E_e)_{(\rm NLO)} - \frac{E_e}{m_N}\,
\bar{H}(E_e)_{(\rm NLO)} - \frac{E_e}{m_N}\, \bar{K}_e(E_e)_{(\rm
  NLO)}\Big(1 - \frac{m_e}{E_e}\Big) = \nonumber\\
\hspace{-0.3in}&& =1.97 \times 10^{-5}\,\frac{E_e}{E_0} - 5.49
\times 10^{-5}\, \frac{E^2_e}{E^2_0},
\nonumber\\
\hspace{-0.3in}&& \bar{H}^{(4)}(E_e)_{(\rm N^2LO)} = 3\,
\frac{E_e}{m_N}\, \bar{H}(E_e)_{(\rm NLO)}\quad,\quad
\bar{N}^{(4)}(E_e)_{(\rm N^2LO)} = 3\, \frac{E_e}{m_N}\, \bar{N}(E_e)_{(\rm NLO)},
\nonumber\\
 \hspace{-0.21in}&& \bar{Q}^{(4)}_e(E_e)_{(\rm N^2LO)} = 3\,
 \frac{E_e}{m_N} \, \bar{Q}_e(E_e)_{(\rm NLO)} =  1.96 \times 10^{-5}\,
 \frac{E^2_e}{E^2_0},\nonumber\\
 \hspace{-0.3in}&& \bar{K}^{(4)}_e(E_e)_{(\rm N^2LO)} = 3\,
 \frac{E_e}{m_N}\, \bar{K}_e(E_e)_{(\rm NLO)} - 3\, \frac{E_e}{m_N}\,
 \bar{G}(E_e)_{(\rm NLO)}\Big(1 + \frac{m_e}{E_e}\Big) = 
 8.30 \times 10^{-5}\,\frac{E^2_e}{E^2_0}.
\end{eqnarray}
We have given the corrections $\zeta^{(4)}(E_e)_{(\rm N^2LO)}$ and
$\bar{X}^{(4)}(E_e)_{(\rm N^2L0)}$ in the form of polynomials in the
variable $E_e/E_0$. The coefficients of the polynomials are calculated
at the neglect of the contributions of order of a few parts of
$10^{-6}$. The same order of magnitude possess the corrections
$\bar{H}^{(4)}(E_e)_{(\rm N^2LO)}$ and $\bar{N}^{(4)}(E_e)_{(\rm
  N^2LO)}$, respectively.

\subsection*{\bf 5. Corrections $\bar{X}(E_e)_{(\rm LO)}
  \zeta(E_e)_{(\rm N^2LO)}$}

For the corrections $\bar{X}(E_e)_{(\rm LO)} \zeta(E_e)_{(\rm N^2LO)}$
we obtain the following analytical expressions
\begin{eqnarray}\label{eq:17}
\hspace{-0.15in}\bar{a}(E_e)_{(\rm LO)}\zeta(E_e)_{(\rm N^2LO)} &=&
\frac{1 - g^2_A}{1 + 3 g^2_A}\,\zeta(E_e)_{(\rm N^2LO)}\;,\;
\bar{A}(E_e)_{(\rm LO)} \zeta(E_e)_{(\rm N^2LO)} = 2\, \frac{g_A (1 -
  g_A)}{1 + 3 g^2_A}\, \zeta(E_e)_{(\rm N^2LO)},\nonumber\\
\hspace{-0.21in}\bar{B}(E_e)_{(\rm LO)} \zeta(E_e)_{(\rm N^2LO)} &=&
2\, \frac{g_A (1 + g_A)}{1 + 3 g^2_A}\, \zeta(E_e)_{(\rm
  N^2LO)} \;,\; \bar{K}_n(E_e)_{(\rm LO)} \zeta(E_e)_{(\rm
  N^2LO)} = \bar{Q}_n(E_e)_{(\rm LO)} \zeta(E_e)_{(\rm N^2LO)} =
0,\nonumber\\
\hspace{-0.21in} \bar{G}(E_e)_{(\rm LO)} \zeta(E_e)_{(\rm N^2LO)} &=&
- \zeta(E_e)_{(\rm N^2LO)} \;,\; \bar{H}(E_e)_{(\rm LO)} \zeta(E_e)_{(\rm N^2LO)} =
- \frac{1 - g^2_A}{1 + 3 g^2_A}\, \frac{m_e}{E_e}\,\zeta(E_e)_{(\rm
  N^2LO)}, \nonumber\\
\hspace{-0.21in}\bar{N}(E_e)_{(\rm LO)} \zeta(E_e)_{(\rm N^2LO)} &=& -
2 \,\frac{g_A(1 - g_A)}{1 + 3 g^2_A}\, \frac{m_e}{E_e}\,
\zeta(E_e)_{(\rm N^2LO)} \;,\; \bar{Q}_e(E_e)_{(\rm LO)}
\zeta(E_e)_{(\rm N^2LO)} = - 2\, \frac{g_A(1 - g_A)}{1 + 3 g^2_A}\,
\zeta(E_e)_{(\rm N^2LO)},\nonumber\\
 \hspace{-0.21in}\bar{K}_e(E_e)_{(\rm LO)} \zeta(E_e)_{(\rm N^2LO)}
 &=& - \frac{1 - g^2_A}{1 + 3 g^2_A}\, \zeta(E_e)_{(\rm N^2LO)},
\end{eqnarray}
where we have used the results obtained in \cite{Ivanov2013,
  Ivanov2017, Ivanov2018, Ivanov2019}.  According to our analysis,
carried out above, the correction $\zeta(E_e)_{(\rm N^2LO)}$ contains
contributions of order of  $10^{-5}$ from $\zeta^{(2)}(E_e)_{(\rm N^2LO)}$
and $\zeta^{(4)}(E_e)_{(\rm N^2LO)}$ only. So we may set
\begin{eqnarray}\label{eq:18}
\zeta(E_e)_{(\rm N^2LO)} &=& \zeta^{(2)}(E_e)_{(\rm N^2LO)} +
\zeta^{(4)}(E_e)_{(\rm N^2LO)} = - 1.97 \times
10^{-5}\,\frac{E_e}{E_0} + 7.09 \times 10^{-5}\, \frac{E^2_e}{E^2_0}.
\end{eqnarray}
Because of the factors $g_A(1 - g_A)/(1 + 3 g^2_A) = - 0.060$ and $(1
- g^2_A)/(1 + 3 g^2_A) = - 0.107$ all corrections, proportional to
these factors, are of order of a few parts of $10^{-6}$ or even
smaller. As a result, between the corrections in Eq.(\ref{eq:17}) we
may take into account the corrections $\bar{B}(E_e)_{(\rm LO)}
\zeta(E_e)_{(\rm N^2LO)}$ and $\bar{G}(E_e)_{(\rm LO)}
\zeta(E_e)_{(\rm N^2LO)}$, where $\bar{B}(E_e)_{(\rm LO)} $ and
$\bar{G}(E_e)_{(\rm LO)}$ are of order $O(1)$. We get
\begin{eqnarray}\label{eq:19}
 \bar{B}(E_e)_{(\rm LO)} \zeta(E_e)_{(\rm N^2LO)} &=& - 1.94 \times
 10^{-5}\,\frac{E_e}{E_0} + 7.00 \times 10^{-5}\,
 \frac{E^2_e}{E^2_0},\nonumber\\ \bar{G}(E_e)_{(\rm LO)}
 \zeta(E_e)_{(\rm N^2LO)} &=& 1.97 \times 10^{-5}\,\frac{E_e}{E_0} -
 7.09 \times 10^{-5}\, \frac{E^2_e}{E^2_0}.
\end{eqnarray}
Thus, the corrections $\bar{B}(E_e)_{(\rm LO)} \zeta(E_e)_{(\rm
  N^2LO)}$ and $\bar{G}(E_e)_{(\rm LO)} \zeta(E_e)_{(\rm N^2LO)}$
possess a required order of a few parts of $10^{-5}$ only.

\subsection*{\bf 6. Corrections $\bar{X}(E_e)_{(\rm LO)}
  \zeta^2(E_e)_{(\rm NLO)}$}

The corrections $\bar{X}(E_e)_{(\rm LO)} \zeta^2(E_e)_{(\rm NLO)}$ we
calculate by using the results obtained in \cite{Ivanov2013,
  Ivanov2017, Ivanov2018, Ivanov2019}. The correction
$\zeta(E_e)_{(\rm NLO)}$ is adduced also in Appendix C. We get
\begin{eqnarray}\label{eq:20}
\hspace{-0.15in}\bar{a}(E_e)_{(\rm LO)}\zeta^2(E_e)_{(\rm NLO)} &=&
\frac{1 - g^2_A}{1 + 3 g^2_A}\,\zeta^2(E_e)_{(\rm NLO)} = - 1.05\times
10^{-5}\,\frac{E^2_e}{E^2_0}, \nonumber\\
\hspace{-0.21in}\bar{A}(E_e)_{(\rm LO)} \zeta^2(E_e)_{(\rm NLO)} &=&
2\, \frac{g_A (1 - g_A)}{1 + 3 g^2_A}\, \zeta^2(E_e)_{(\rm NLO)} = -
1.17\times 10^{-5}\,\frac{E^2_e}{E^2_0},\nonumber\\
\hspace{-0.21in}\bar{B}(E_e)_{(\rm LO)} \zeta^2(E_e)_{(\rm NLO)} &=&
2\, \frac{g_A (1 + g_A)}{1 + 3 g^2_A}\, \zeta^2(E_e)_{(\rm NLO)} =
- 6.98 \times 10^{-5}\, \frac{E_e}{E_0} + 9.67
\times 10^{-5}\, \frac{E^2_e}{E^2_0},\nonumber\\
\hspace{-0.21in}\bar{K}_n(E_e)_{(\rm LO)} \zeta^2(E_e)_{(\rm
  NLO)} &=& \bar{Q}_n(E_e)_{(\rm LO)} \zeta^2(E_e)_{(\rm NLO)} =
0,\nonumber\\
\hspace{-0.21in} \bar{G}(E_e)_{(\rm LO)} \zeta^2(E_e)_{(\rm NLO)} &=&
- \zeta^2(E_e)_{(\rm NLO)} = 7.07 \times 10^{-5}\,\frac{E_e}{E_0} -
9.79 \times 10^{-5}\, \frac{E^2_e}{E^2_0},\nonumber\\
\hspace{-0.21in}\bar{H}(E_e)_{(\rm LO)} \zeta^2(E_e)_{(\rm NLO)} &=&
- \frac{1 - g^2_A}{1 + 3 g^2_A}\, \frac{m_e}{E_e}\,\zeta^2(E_e)_{(\rm
  NLO)}, \nonumber\\
\hspace{-0.21in}\bar{N}(E_e)_{(\rm LO)} \zeta^2(E_e)_{(\rm NLO)} &=& -
2 \,\frac{g_A(1 - g_A)}{1 + 3 g^2_A}\, \frac{m_e}{E_e}\,
\zeta^2(E_e)_{(\rm NLO)}, \nonumber\\
\hspace{-0.21in} \bar{Q}_e(E_e)_{(\rm LO)} \zeta^2(E_e)_{(\rm NLO)}
&=& - 2\, \frac{g_A(1 - g_A)}{1 + 3 g^2_A}\, \zeta^2(E_e)_{(\rm NLO)}
= 1.17 \times 10^{-5}\, \frac{E^2_e}{E^2_0},\nonumber\\
 \hspace{-0.21in}\bar{K}_e(E_e)_{(\rm LO)} \zeta^2(E_e)_{(\rm NLO)}
 &=& - \frac{1 - g^2_A}{1 + 3 g^2_A}\, \zeta^2(E_e)_{(\rm NLO)} = 1.05
 \times 10^{-5}\, \frac{E^2_e}{E^2_0},
\end{eqnarray}
where we have neglected the contributions of order of a few parts of
$10^{-6}$. The same order of magnitude possess the corrections
$\bar{H}(E_e)_{(\rm LO)} \zeta^2(E_e)_{(\rm NLO)}$ and
$\bar{N}(E_e)_{(\rm LO)} \zeta^2(E_e)_{(\rm NLO)}$, respectively, the
contributions of which we have neglected.

\subsection*{\bf 7. Corrections $\bar{X}(E_e)_{(\rm NLO)}
  \zeta(E_e)_{(\rm NLO)}$}

The calculation of corrections $\bar{X}(E_e)_{(\rm NLO)}
\zeta(E_e)_{(\rm NLO)}$ we may perform by using the results obtained
in \cite{Ivanov2013, Ivanov2017, Ivanov2018, Ivanov2019}, which we
have adduced in Appendix C. Since the analytical expressions of these
corrections are rather bulky, we give them in the from of polynomials
in the variable $E_e/E_0$ only. We get
\begin{eqnarray}\label{eq:21}
\bar{a}(E_e)_{(\rm NLO)}\zeta(E_e)_{(\rm NLO)} &=& 7.17 \times
10^{-5}\,\frac{E_e}{E_0} - 1.01 \times 10^{-4}
\,\frac{E^2_e}{E^2_0},\nonumber\\ \bar{A}(E_e)_{(\rm
  NLO)}\zeta(E_e)_{(\rm NLO)} &=& 1.43 \times 10^{-5}\,\frac{E_e}{E_0}
- 2.90 \times
10^{-5}\,\frac{E^2_e}{E^2_0},\nonumber\\ \bar{B}(E_e)_{(\rm
  NLO)}\zeta(E_e)_{(\rm NLO)} &=&  - 6.92 \times
10^{-5}\,\frac{E_e}{E_0} + 9.37 \times 10^{-5}\,
\frac{E^2_e}{E^2_0},\nonumber\\ \bar{Q}_n(E_e)_{(\rm
  NLO)}\zeta(E_e)_{(\rm NLO)} &=&  5.74 \times
10^{-5}\,\frac{E_e}{E_0} - 7.19 \times
10^{-5}\,\frac{E^2_e}{E^2_0},\nonumber\\ \bar{G}(E_e)_{(\rm
  NLO)}\zeta(E_e)_{(\rm NLO)} &=&  7.07 \times
10^{-5}\,\frac{E_e}{E_0} - 9.79 \times 10^{-5}\, \frac{E^2_e}{E^2_0},
\nonumber\\ \bar{H}(E_e)_{(\rm NLO)}\zeta(E_e)_{(\rm NLO)} &=&- 1.94
\times 10^{-5} + 1.51 \times
10^{-5}\,\frac{E_e}{E_0},\nonumber\\ \bar{Q}_e(E_e)_{(\rm
  NLO)}\zeta(E_e)_{(\rm NLO)} &=& - 1.44\times 10^{-5}\,
\frac{E_e}{E_0} + 4.70 \times
10^{-5}\,\frac{E^2_e}{E^2_0},\nonumber\\ \bar{K}_e(E_e)_{(\rm
  NLO)}\zeta(E_e)_{(\rm NLO)} &=& - 4.70 \times
10^{-5}\,\frac{E_e}{E_0} + 1.01 \times 10^{-4}\,\frac{E^2_e}{E^2_0}.
\end{eqnarray}
The corrections $\bar{K}_n(E_e)_{(\rm NLO)}\zeta(E_e)_{(\rm NLO)}$ and
$\bar{N}(E_e)_{(\rm NLO)}\zeta(E_e)_{(\rm NLO)}$ are of order of a few
parts of $10^{-6}$.

\section{Corrections $\zeta(E_e)_{(\rm N^2LO)}$ and $X(E_e)_{(\rm N^2LO)}$ in
  the form of polynomials in the variable $E_e/E_0$}
\label{sec:barX2}

Summing up the contributions of order of  $10^{-5}$ we obtain the
corrections $\zeta(E_e)_{(\rm N^2LO)}$ and $X(E_e)_{(\rm N^2LO)}$,
caused by weak magnetism and proton recoil calculated to
next-to-next-to-leading order in the large nucleon mass $m_N$
expansion. We give them in the form of polynomials in the variable
$E_e/E_0$:
\begin{eqnarray*}
\hspace{-0.3in} \zeta(E_e)_{(\rm N^2LO)} &=& \zeta^{(2)}(E_e)_{(\rm
  N^2LO)} + \zeta^{(4)}(E_e)_{(\rm N^2LO)} + \zeta(E_e)_{\rm FF} = -
3.27 \times 10^{-5}\,\frac{E_e}{E_0} + 8.39 \times 10^{-5}\,
\frac{E^2_e}{E^2_0},\nonumber\\ 
\hspace{-0.3in} a(E_e)_{(\rm N^2LO)} &=& \bar{a}^{(2)}(E_e)_{(\rm
  N^2LO)} + \bar{a}^{(4)}(E_e)_{(\rm N^2LO)} + \bar{a}(E_e)_{(\rm LO)}
\zeta^2(E_e)_{(\rm NLO)} - \bar{a}(E_e)_{(\rm NLO)} \zeta(E_e)_{(\rm
  NLO)}\nonumber\\ &+& a(E_e)_{\rm FF} = - 3.90
\times 10^{-5}\, \frac{E_e}{E_0}, \nonumber\\
\hspace{-0.3in} A(E_e)_{(\rm N^2LO)} &=& \bar{A}^{(4)}(E_e)_{(\rm N^2LO)}
+ \bar{A}(E_e)_{(\rm LO)} \zeta^2(E_e)_{(\rm NLO)} -
\bar{A}(E_e)_{(\rm NLO)} \zeta(E_e)_{(\rm NLO)} = \nonumber\\ &=& -
1.43 \times 10^{-5}\, \frac{E_e}{E_0}, \nonumber\\
\hspace{-0.3in}B(E_e)_{(\rm N^2LO)} &=& \bar{B}^{(2)}(E_e)_{(\rm
  N^2LO)} + \bar{B}^{(4)}(E_e)_{(\rm N^2LO)} - \bar{B}(E_e)_{(\rm LO)}
\zeta(E_e)_{(\rm N^2LO)} + \bar{B}(E_e)_{(\rm LO)} \zeta^2(E_e)_{(\rm
  NLO)} \nonumber\\ &-& \bar{B}(E_e)_{(\rm NLO)} \zeta(E_e)_{(\rm
  NLO)} + B(E_e)_{\rm FF}= - 1.62 \times 10^{-5}\,
\frac{E^2_e}{E^2_0}, \nonumber\\
\hspace{-0.3in}K_n(E_e)_{(\rm N^2LO)} &=& \bar{K}^{(4)}_n(E_e)_{(\rm
  N^2LO)} = 1.51 \times 10^{-5}\, \frac{E^2_e}{E^2_0},
\nonumber\\
\hspace{-0.3in}Q_n(E_e)_{(\rm N^2LO)} &=& \bar{Q}^{(2)}_n(E_e)_{(\rm
  N^2LO)} + \bar{Q}^{(4)}_n(E_e)_{(\rm N^2LO)} - \bar{Q}_n(E_e)_{(\rm
  NLO)} \zeta(E_e)_{(\rm NLO)} = \nonumber\\ &=&  -
2.95 \times 10^{-5}\, \frac{E_e}{E_0} - 1.97 \times 10^{-5}\,
\frac{E^2_e}{E^2_0},\nonumber\\
\hspace{-0.3in} G(E_e)_{(\rm N^2LO)} &=& \bar{G}^{(2)}(E_e)_{(\rm
  N^2LO)} + \bar{G}^{(4)}(E_e)_{(\rm N^2LO)} - \bar{G}(E_e)_{(\rm LO)}
\zeta(E_e)_{(\rm N^2LO)} + \bar{G}(E_e)_{(\rm LO)} \zeta^2(E_e)_{(\rm
  NLO)} \nonumber\\ &-& \bar{G}(E_e)_{(\rm NLO)} \zeta(E_e)_{(\rm
  NLO)} + G(E_e)_{\rm FF}= 0, \nonumber\\
\hspace{-0.3in} H(E_e)_{(\rm N^2LO)} &=& - \bar{H}(E_e)_{(\rm NLO)}
 \zeta(E_e)_{(\rm NLO)} = 1.94 \times 10^{-5} - 1.51 \times 10^{-5}\,
 \frac{E_e}{E_0},\nonumber\\
\hspace{-0.3in} N(E_e)_{(\rm N^2LO)} &=& -
\bar{N}(E_e)_{(\rm NLO)} \zeta(E_e)_{(\rm NLO)} = 0,\nonumber\\
   \end{eqnarray*}
\begin{eqnarray}\label{eq:22}
\hspace{-0.3in} Q_e(E_e)_{(\rm N^2LO)} &=&
 \bar{Q}^{(4)}_e(E_e)_{(\rm N^2LO)} + \bar{Q}_e(E_e)_{(\rm LO)}
 \zeta^2(E_e)_{(\rm NLO)} - \bar{Q}_e(E_e)_{(\rm NLO)}
 \zeta(E_e)_{(\rm NLO)} = \nonumber\\ &=& 1.44 \times 10^{-5}\,
 \frac{E_e}{E_0} - 1.57 \times 10^{-5}\,
 \frac{E^2_e}{E^2_0},\nonumber\\
\hspace{-0.3in} K_e(E_e)_{(\rm N^2LO)} &=& \bar{K}^{(2)}_e(E_e)_{(\rm
  N^2LO)} + \bar{K}^{(4)}_e(E_e)_{(\rm N^2LO)} + \bar{K}_e(E_e)_{(\rm
  LO)} \zeta^2(E_e)_{(\rm NLO)} - \bar{K}_e(E_e)_{(\rm NLO)}
\zeta(E_e)_{(\rm NLO)}\nonumber\\
\hspace{-0.3in}&+& K_e(E_e)_{\rm FF} = 3.40 \times
10^{-5}\,\frac{E_e}{E_0}+ 2.95 \times 10^{-5}\, \frac{E^2_e}{E^2_0},
\end{eqnarray}
where we have added the corrections $X(E_e)_{\rm FF}$ for $X = \zeta,
a, B, G$ and $K_e$, induced by the isovector and axial-vector form
factors of the neutron taken in the linear approximation for the
square four-momentum transfer (see Appendix A).  The corrections,
given in Eq.(\ref{eq:22}), illustrate an existence of corrections
$O(E^2_e/m^2_N)$ of order of a few parts of $10^{-5}$, caused by weak
magnetism and proton recoil and calculated to next-to-next-to-leading
order in the large nucleon mass $m_N$ expansion. These corrections,
calculated with a theoretical accuracy of about a few parts of
$10^{-6}$, are given in the form, which can be, in principle, applied
to the description of the SM theoretical background of the neutron
beta decay at the level of a few parts of $10^{-5}$.

\section{Conclusion}
\label{sec:Schluss}

We have calculated a complete set of the model--independent
next-to-next-to-leading order corrections $O(E^2_e/m^2_N)$ in the
large nucleon mass $m_N$ expansion, caused by weak magnetism and
proton recoil, to the neutron lifetime and correlation coefficients of
the neutron beta decay for a polarized neutron, a polarized electron
and an unpolarized proton.  We have given a detailed analysis of the
structure of these corrections. In addition we have added corrections,
caused by the isovector and axial--vector form factors of the nucleon,
calculated to linear approximation in the square of four-momentum
transfer with the form factors taken in the dipole approximation. The
notation ``large nucleon mass $m_N$ expansion'' means that we make an
expansion in powers of $1/m_N$, were the nucleon mass is finite but
much larger than the momentum of the proton and energies and momenta
of decay leptons. The effect of the finite nucleon mass appears only
to next-to-leading and higher orders in the large nucleon mass $m_N$
expansion or in the terms proportional to powers of $1/m_N$. For the
calculation of the corrections $O(E^2_e/m^2_N)$ we have used a
standard technique \cite{Bilenky1959, Gudkov2006, Ivanov2013,
  Ivanov2017, Ivanov2018, Ivanov2019} which is well expounded in
Appendix of Ref. \cite{Ivanov2013}.  We have presented the corrections
$O(E^2_e/m^2_N)$ in the form of polynomials in the variable $E_e/E_0$
with coefficients of order of $10^{-5}$ and even $10^{-4}$, calculated
at the neglect of the terms of order of a few parts of $10^{-6}$.
Together with Wilkinson's corrections of order of $10^{-5}$
\cite{Wilkinson1982} (see also \cite{Ivanov2013, Ivanov2017,
  Ivanov2018, Ivanov2019}) and radiative corrections of order
$O(\alpha E_e/m_N) \sim 10^{-5}$ \cite{Ivanov2019a, Ivanov2020a} the
corrections of order $O(E_e/m^2_N) \sim 10^{-5}$ define the SM
theoretical background of the analysis of the neutron beta decay at
the level of a few parts of $10^{-5}$. Such a set of corrections
should provide an improved level of theoretical investigations of the
neutron beta decay within the SM.  The representation of the
corrections of order $O(E^2_e/m^2_N) \sim 10^{- 5}$ in the form of
polynomials in the variable $E_e/E_0$ makes them easily applicable for
the analysis of experimental data on asymmetries of the neutron beta
decay for searches of traces of interactions beyond the SM at the
level of a few parts of $10^{-5}$. For example, in the experimental
electron-energy region $0.811\,{\rm MeV} \le E_e \le 1.211\,{\rm MeV}$
\cite{Abele2018} and $0.708\,{\rm MeV} \le E_e \le 1.205\,{\rm MeV}$
\cite{Abele2019} the correction $A^{(W)}(E_e)_{(\rm N^2LO)}$ to the
correlation coefficient $A^{(W)}(E_e) = A(E_e) +
\frac{1}{3}\,Q_n(E_e)$xs, defining the electron asymmetry in the
neutron beta decay \cite{Wilkinson1982, Abele2018, Abele2019} (see
also \cite{Ivanov2013}), varies in the limits $- 1.77 \times 10^{-5}
\ge A^{(W)}(E_e)_{(\rm N^2LO)} \ge - 2.84 \times 10^{-5}$ and $- 1.52
\times 10^{-5} \ge A^{(W)}(E_e)_{(\rm N^2LO)} \ge - 2.82\times
10^{-5}$, respectively.  Recently \cite{Abele2019} the experimental
data on the measurements of this asymmetry have been analyzed for a
search of the contribution of the Fierz interference term $b$. The
theoretical expression for the correlation coefficient $A^{(W)}(E_e)$
is calculated in the SM at the level of $10^{-3}$ \cite{Wilkinson1982}
(see also \cite{Ivanov2013}), whereas the experimental uncertainties
of the electron asymmetry are at the level of a few parts of
$10^{-4}$, namely, $A^{(W)} = - 0.11985 \pm 0.00021$ \cite{Abele2018}
and $A^{(W)} = - 0.11972 \pm 0.00025$ \cite{Abele2019},
respectively. This implies that the experimental analysis of the
neutron beta decay is more precise than the theoretical one. As has
been pointed out in \cite{Abele2019}, for such uncertainties the
experimental data on the electron asymmetry are consistent with the
Fierz interference term $b = 0$ . Formally, non-zero values for the
Fierz interference term $b$, varying in the limits $- 0.018 \ge b \ge
0.052$, can be extracted from the experimental data on the electron
asymmetry for experimental uncertainties of a few parts of $10^{-3}$,
namely, $A^{(W)} = - 0.1209 \pm 0.0015$. Such an uncertainty is
commensurable with the level of the SM theoretical calculation of the
correlation coefficient $A^{(W)}(E_e)$ \cite{Wilkinson1982,
  Ivanov2013}. So for an improvement of experimental constraints on
the Fierz interference term by measuring the electron asymmetry of the
neutron beta decay it is desirable to improve the theoretical level of
the SM calculation of the electron asymmetry from $10^{-3}$ to
$10^{-5}$. Such an improvement should make meaningful an improvement
of experimental uncertainties by order of magnitude, i.e. from a few
parts of $10^{-4}$ to a few parts of $10^{-5}$.  For the SM definition
of the electron asymmetry at the level of $10^{-5}$ one may use
Wilkinson's corrections \cite{Wilkinson1982} (see also
\cite{Ivanov2013, Ivanov2017, Ivanov2018, Ivanov2019}), the radiative
corrections $O(\alpha E_e/m_N) \sim 10^{-5}$ \cite{Ivanov2019a,
  Ivanov2020a}, calculated as next-to-leading order corrections in the
large nucleon mass $m_N$ expansion to Sirlin's radiative corrections
of order $O(\alpha/\pi)$ \cite{Sirlin1967}, and the model-independent
corrections $O(E^2_e/m^2_N) \sim 10^{-5}$. At the new level of
theoretical and experimental analysis of the electron asymmetry of the
neutron beta decay one might expect new much better constraints on the
Fierz interference term than $ - 0.018 \le b \le 0.052$ reported in
\cite{Abele2019}. Then, we would like to mention that, according to
Paul \cite{Paul2009}, the desired accuracy of experimental
investigations of the neutron lifetime should be $\delta \tau_n <
0.08\,{\rm s}$, i.e. at the level of sensitivity of a few parts of
$10^{-5}$. We are planning to calculate the neutron lifetime and
correlation coefficients of the neutron beta decay at the level of a
few parts of $10^{-5}$ in our forthcoming publication
\cite{Ivanov2021}.

\section{Acknowledgements}

We thank Hartmut Abele for discussions stimulating the work under
corrections of order of  $10^{-5}$ to the neutron lifetime and correlation
coefficients of the neutron $\beta^-$--decays with different
polarization states of the neutron and massive decay fermions. The
work of A. N. Ivanov was supported by the Austrian ``Fonds zur
F\"orderung der Wissenschaftlichen Forschung'' (FWF) under contracts
P31702-N27 and P26636-N20, and ``Deutsche F\"orderungsgemeinschaft''
(DFG) AB 128/5-2. The work of R. H\"ollwieser was supported by the
Deutsche Forschungsgemeinschaft in the SFB/TR 55. The work of
M. Wellenzohn was supported by the MA 23.

\newpage

\section*{Appendix A: The amplitude of the neutron beta decay with
  weak magnetism and proton recoil to order $O(E^2_e/m^2_N)$ }
\renewcommand{\theequation}{A-\arabic{equation}}
\setcounter{equation}{0}

Following \cite{Ivanov2013} the amplitude of the neutron beta decay we
rewrite as follows
\begin{eqnarray}\label{eq:A.1} 
M(n \to pe^-\bar{\nu}_e) = -\frac{G_F}{\sqrt{2}}\,V_{ud}\,{\cal M}_n,
\end{eqnarray}
where ${\cal M}_n = [\bar{u}_p O_{\mu} u_n] [\bar{u}_e\gamma^{\mu}(1 -
  \gamma^5)v_{\bar{\nu}}]$ and the matrix
$O_{\mu}$ takes the form
\begin{eqnarray}\label{eq:A.2} 
O_{\mu} = \gamma_{\mu}(1 - g_A \gamma^5) + i\,\frac{\kappa}{2
  m_N}\,\sigma_{\mu\nu}(k_p - k_n)^{\nu},
\end{eqnarray}
where $m_N = (m_n + m_p)/2$ is the nucleon mass.  In terms of the time
and space components of the matrix $O_{\mu} = (O^0, - \vec{O}\,)$ the
amplitude ${\cal M}_n$ is defined by \cite{Ivanov2013}
\begin{eqnarray}\label{eq:A.3} 
{\cal M}_n = [\bar{u}_p O^0 u_n][\bar{u}_e\gamma^0(1 -
\gamma^5)v_{\bar{\nu}}] - [\bar{u}_p \vec{O}
u_n]\cdot [\bar{u}_e \vec{\gamma}\,(1 - \gamma^5)v_{\bar{\nu}}].
\end{eqnarray}
The time $O^0$ and spacial $\vec{O}$ components of the matrix $O_{\mu}
= (O^0, - \vec{O}\,)$ we determine in the large nucleon mass $m_N$
expansion keeping the terms proportional to $1/m^2_N$. We get
\begin{eqnarray}\label{eq:A.4}
 \hspace{-0.3in} O^0 =\left(\begin{array}{ccc} 1 & {\displaystyle -
     g_A + \kappa\, \frac{\vec{\sigma}\cdot \vec{k}_p}{2
       m_N}}\\ {\displaystyle g_A + \kappa\, \frac{\vec{\sigma}\cdot
       \vec{k}_p}{2 m_N}} & - 1\\
    \end{array}\right)
\end{eqnarray}
and 
\begin{eqnarray}\label{eq:A.5}
 \hspace{-0.3in} \vec{O} = \left(\begin{array}{ccc} {\displaystyle -
     g_A \vec{\sigma} + i \kappa \,\frac{\vec{\sigma}\times
       \vec{k}_p}{2 m_N}} & {\displaystyle \vec{\sigma}\,\Big(1 -
     \kappa\, \frac{E_0}{2 m_N} - \kappa\, \frac{E^2_0 - m^2_e -
       \vec{k}^{\,2}_p}{4 m^2_N}\Big)}\\ {\displaystyle
     -\,\vec{\sigma}\,\Big(1 + \kappa\, \frac{E_0}{2 m_N} + \kappa\,
     \frac{E^2_0 - m^2_e - \vec{k}^{\,2}_p}{4 m^2_N}\Big)} &
   {\displaystyle g_A \vec{\sigma}+ i \kappa\,
     \frac{\vec{\sigma}\times \vec{k}_p}{2 m_N}} \\
    \end{array}\right),
\end{eqnarray}
where $E_0 = (m^2_n - m^2_p + m^2_e)/2 m_n = 1.2926\, {\rm MeV}$ is
the end--point energy of the electron--energy spectrum of the neutron
beta decay \cite{Ivanov2013}, $\vec{k}_p = - \vec{k}_e -
\vec{k}_{\bar{\nu}}$ is 3--momentum transferred momentum in terms of
the 3--momenta of the electron $\vec{k}_e$ and antineutrino
$\vec{k}_{\bar{\nu}}$, $\vec{k}^{\,2}_p = E^2_e - m^2_e +
E^2_{\bar{\nu}} + 2 \vec{k}_e \cdot \vec{k}_{\bar{\nu}}$.

For the calculation of the amplitude of the neutron beta decay we use
the Dirac wave functions of the neutron and the proton in the momentum
representation
\begin{eqnarray}\label{eq:A.6}
 \hspace{-0.3in} u_n(\vec{0},\sigma_n) = \sqrt{2
   m_n}\Big(\begin{array}{c}\varphi_n \\ 0
 \end{array}\Big) \quad,\quad u_p(\vec{k}_p,\sigma_p) = \sqrt{E_p +
 m_p}\left(\begin{array}{c}\varphi_p \\ {\displaystyle
 \frac{\vec{\sigma}\cdot \vec{k}_p}{E_p + m_p}\,\varphi_p }
 \end{array}\right),
\end{eqnarray}
where the Pauli spinorial wave functions $\varphi_n$ and $\varphi_p$
depend on the polarisations $\sigma_n = \pm 1/2$ and $\sigma_p = \pm
1/2$, respectively.  For the calculation of the matrix elements
$[\bar{u}_p O^0 u_n]$ and $[\bar{u}_p \vec{O} u_n]$ we use the
following expansions
\begin{eqnarray}\label{eq:A.7}
 \hspace{-0.3in}\sqrt{2 m_n(E_p + m_p)} = 2m_n \Big(1 - \frac{E_0}{2
   m_N} - \frac{E^2_0 - 2 m^2_e - \vec{k}^{\,2}_p }{8 m^2_N}\Big),
 \quad,\quad \frac{\vec{\sigma}\cdot \vec{k}_p}{E_p + m_p} =
 \frac{\vec{\sigma}\cdot \vec{k}_p}{2 m_N} +
 \frac{E_0(\vec{\sigma}\cdot \vec{k}_p)}{4 m^2_N}.
\end{eqnarray}
As a result, we get
\begin{eqnarray}\label{eq:A.8}
 \hspace{-0.3in}&&[\bar{u}_p O^0 u_n] = 2m_n\Big\{\Big(1 -
 \frac{E_0}{2m_N} - \frac{E^2_0 - 2 m^2_e + (2 \kappa -
   1)\,\vec{k}^{\,2}_p}{8 m^2_N}\Big)\, [\varphi^{\dagger}_p
   \varphi_n] - \frac{g_A}{2 m_N}\, [\varphi^{\dagger}_p
   (\vec{\sigma}\cdot \vec{k}_p) \varphi_n]\Big\}
\end{eqnarray}
and
\begin{eqnarray}\label{eq:A.9}
 \hspace{-0.3in}[\bar{u}_p \vec{O} u_n] &=& 2m_n \,\Big\{ - g_A \Big(1
 - \frac{E_0}{2m_N} - \frac{E^2_0 - 2 m^2_e - \vec{k}^{\,2}_p}{8
   m^2_N}\Big)\, [\varphi^{\dagger}_p\vec{\sigma}\,\varphi_n] +
 i\,\frac{\kappa }{2 m_N}\,\Big(1 - \frac{E_0}{2m_N}\Big)\,
 [\varphi^{\dagger}_p(\vec{\sigma}\times \vec{k}_p)\varphi_n]\nonumber\\
 \hspace{-0.3in}&+& \frac{1}{2 m_N}\, \Big(1 + \kappa\,
 \frac{E_0}{2m_N}\Big)[\varphi^{\dagger}_p (\vec{\sigma}\cdot
   \vec{k}_p) \vec{\sigma}\, \varphi_n]\Big\}.
\end{eqnarray}
 Using the relation $(\vec{\sigma}\cdot \vec{k}_p)\vec{\sigma} =
 \vec{k}_p + i\,(\vec{\sigma}\times \vec{k}_p)$ we rewrite the
 r.h.s. of Eq.(\ref{eq:A.9}) as follows
\begin{eqnarray}\label{eq:A.10}
 \hspace{-0.3in}[\bar{u}_p \vec{O} u_n] &=& 2m_n \,\Big\{ - g_A \Big(1
 - \frac{E_0}{2m_N} - \frac{E^2_0 - 2 m^2_e - \vec{k}^{\,2}_p}{8
   m^2_N}\Big)\, [\varphi^{\dagger}_p\vec{\sigma}\,\varphi_n] +
 i\,\frac{\kappa + 1}{2 m_N}\, [\varphi^{\dagger}_p(\vec{\sigma}\times
   \vec{k}_p)\varphi_n]\nonumber\\
 \hspace{-0.3in}&+& \frac{\vec{k}_p}{2 m_N}\, \Big(1 + \kappa\,
 \frac{E_0}{2m_N}\Big)[\varphi^{\dagger}_p \varphi_n]\Big\}.
\end{eqnarray}
For the amplitude ${\cal M}_n$ we obtain the following expression
\begin{eqnarray}\label{eq:A.11}
\hspace{-0.3in}&&{\cal M}_n = 2 m_n \Big\{\Big(1 - \frac{E_0}{2m_N} -
\frac{E^2_0 - 2 m^2_e + (2 \kappa - 1)\,\vec{k}^{\,2}_p}{8
  m^2_N}\Big)\, [\varphi^{\dagger}_p \varphi_n][\bar{u}_e \gamma^0 (1
  - \gamma^5) v_{\bar{\nu}}] + g_A \Big(1 - \frac{E_0}{2m_N} -
\frac{E^2_0 - 2 m^2_e - \vec{k}^{\,2}_p}{8 m^2_N}\Big)\nonumber\\
 \hspace{-0.3in}&& \times \,
        [\varphi^{\dagger}_p\vec{\sigma}\,\varphi_n]\cdot [\bar{u}_e
          \vec{\gamma}\, (1 - \gamma^5) v_{\bar{\nu}}] - \frac{g_A}{2
          m_N}\, [\varphi^{\dagger}_p (\vec{\sigma}\cdot \vec{k}_p)
          \varphi_n][\bar{u}_e \gamma^0 (1 - \gamma^5) v_{\bar{\nu}}]
       - i\,\frac{\kappa + 1}{2
   m_N}\,[\varphi^{\dagger}_p(\vec{\sigma}\times
   \vec{k}_p)\varphi_n]\cdot [\bar{u}_e\vec{\gamma}\,(1 -
   \gamma^5)v_{\bar{\nu}}]  \nonumber\\
 \hspace{-0.3in}&& - \frac{\vec{k}_p}{2 m_N}\, \Big(1 + \kappa\,
 \frac{E_0}{2m_N}\Big)\,[\varphi^{\dagger}_p\varphi_n]\cdot
      [\bar{u}_e\vec{\gamma}\,(1 - \gamma^5)v_{\bar{\nu}}]\Big\}.
\end{eqnarray}
For the transformation of the last term we use the Dirac equations for
the electron and antineutrino. We get
\begin{eqnarray}\label{eq:A.12}
\hspace{-0.3in}- \frac{\vec{k}_p}{2
  m_N}\,[\varphi^{\dagger}_p\varphi_n]\cdot [\bar{u}_e\vec{\gamma}\,(1 -
  \gamma^5)v_{\bar{\nu}}] = \frac{m_n - E_p}{2
  m_N}[\varphi^{\dagger}_p\varphi_n][\bar{u}_e \gamma^0 (1 -
  \gamma^5)v_{\bar{\nu}}] - \frac{m_e}{2
  m_N}[\varphi^{\dagger}_p\varphi_n][\bar{u}_e (1 -
  \gamma^5)v_{\bar{\nu}}],\nonumber\\
 \hspace{-0.3in}&&
\end{eqnarray}
where we have set $E_e + E_{\bar{\nu}} = m_n - E_p$. Making the large
nucleon mass expansion and keeping the contributions of order
$1/m^2_N$ we transcribe the right--hand--side (r.h.s.) of
Eq.(\ref{eq:A.12}) into the form
\begin{eqnarray}\label{eq:A.13}
\hspace{-0.3in}&&- \frac{\vec{k}_p}{2
  m_N}\,[\varphi^{\dagger}_p\varphi_n]\cdot [\bar{u}_e\vec{\gamma}\,(1
  - \gamma^5)v_{\bar{\nu}}] = \Big(\frac{E_0}{2 m_N} + \frac{E^2_0 -
  m^2_e - \vec{k}^{\,2}_p}{4 m^2_N}\Big)
       [\varphi^{\dagger}_p\varphi_n][\bar{u}_e \gamma^0 (1 -
         \gamma^5)v_{\bar{\nu}}] - \frac{m_e}{2
         m_N}[\varphi^{\dagger}_p\varphi_n][\bar{u}_e (1 -
         \gamma^5)v_{\bar{\nu}}].\nonumber\\
 \hspace{-0.3in}&&
\end{eqnarray}
Substituting Eq.(\ref{eq:A.13}) into
Eq.(\ref{eq:A.11}) we get
\begin{eqnarray}\label{eq:A.14}
\hspace{-0.3in}&&{\cal M}_n = 2 m_n\Big\{[\varphi^{\dagger}_p
  \varphi_n][\bar{u}_e \gamma^0 (1 - \gamma^5) v_{\bar{\nu}}] + g_A
\Big(1 - \frac{E_0}{2m_N}\Big)\,
    [\varphi^{\dagger}_p\vec{\sigma}\,\varphi_n]\cdot [\bar{u}_e
      \vec{\gamma}\, (1 - \gamma^5) v_{\bar{\nu}}] - \frac{g_A}{2
      m_N}\, [\varphi^{\dagger}_p (\vec{\sigma}\cdot \vec{k}_p)
      \varphi_n] \nonumber\\
 \hspace{-0.3in}&& \times \, [\bar{u}_e \gamma^0 (1 - \gamma^5)
   v_{\bar{\nu}}] - i\,\frac{\kappa + 1}{2
   m_N}\,[\varphi^{\dagger}_p(\vec{\sigma}\times
   \vec{k}_p)\varphi_n]\cdot [\bar{u}_e\vec{\gamma}\,(1 -
   \gamma^5)v_{\bar{\nu}}] - \frac{m_e}{2 m_N}\,
        [\varphi^{\dagger}_p\varphi_n] [\bar{u}_e (1 -
          \gamma^5)v_{\bar{\nu}}]\Big\} + \delta {\cal M}_n,
\end{eqnarray}
where $\delta {\cal M}_n$ defines the contributions of the terms
proportional to $1/m^2_N$. It is equal to 
\begin{eqnarray}\label{eq:A.15}
\hspace{-0.3in}\delta {\cal M}_n &=& 2 m_n \, \frac{E^2_0}{8 m^2_N}
\Big\{(2 \kappa + 1)\, \frac{q^2}{E^2_0} \,[\varphi^{\dagger}_p
  \varphi_n][\bar{u}_e \gamma^0 (1 - \gamma^5) v_{\bar{\nu}}] - g_A
\frac{q^2 - 2m^2_e}{E^2_0} \,
     [\varphi^{\dagger}_p\vec{\sigma}\,\varphi_n]\cdot [\bar{u}_e
       \vec{\gamma}\, (1 - \gamma^5) v_{\bar{\nu}}] \nonumber\\
 \hspace{-0.3in}&-& 2 \kappa\,
     \frac{m_e}{E_0} \,[\varphi^{\dagger}_p\varphi_n] [\bar{u}_e (1 -
       \gamma^5)v_{\bar{\nu}}]\Big\},
\end{eqnarray}
where $q^2 = E^2_0 - \vec{k}^{\,2}_p = E^2_0 - (\vec{k}_e +
\vec{k}_{\bar{\nu}})^2$. An additional contribution of order
$O(E^2_e/m^2_N)$ to the electron--energy and angular distribution of
the neutron beta decay appears from the phase--volume of the neutron
beta decay. Following \cite{Ivanov2013} we denote the contribution of
the phase--volume of the neutron beta decay as $\Phi_n(k_e, \cos
\theta_{e\bar{\nu}})$, where $\theta_{e\bar{\nu}}$ is an angle between
3--momenta of the electron and antineutrino such as $\cos
\theta_{e\bar{\nu}} = \vec{k}_e \cdot \vec{k}_{\bar{\nu}}$, and define
\begin{eqnarray}\label{eq:A.16}
\hspace{-0.3in}\Phi_n(k_e, \cos \theta_{e\bar{\nu}}) =
\int^{\infty}_0\delta(f(E_{\bar{\nu}}))\,\frac{m_n}{E_p}\,\frac{E^2_{\bar{\nu}}
  dE_{\bar{\nu}}}{(E_0 - E_e)^2},
\end{eqnarray}
where the function $f(E_{\bar{\nu}})$ is given by the energy
conservation in the neutron beta decay $f(E_{\bar{\nu}}) = m_n - E_p -
E_e - E_{\bar{\nu}}$ and $E_p = \sqrt{m^2_p + (\vec{k}_e +
  \vec{k}_{\bar{\nu}})^2} = \sqrt{m^2_p + k^2_e + E^2_{\bar{\nu}} + 2
  k_e E_{\bar{\nu}} \cos\theta_{e\bar{\nu}}}$ is the proton energy
  after the integration over the 3--momentum of the proton, giving
  $\vec{k}_p = - \vec{k}_e - \vec{k}_{\bar{\nu}}$. Using the
  properties of the $\delta$--function the result of the integration
  over the antineutrino energy $E_{\bar{\nu}}$ is equal to
\begin{eqnarray}\label{eq:A.17}
\hspace{-0.3in}\Phi_n(k_e, \cos \theta_{e\bar{\nu}}) =
\frac{m_n}{E_p}\,\frac{E^2_{\bar{\nu}}}{(E_0 -
  E_e)^2}\,\frac{1}{\displaystyle
  \Big|\frac{df(E_{\bar{\nu}})}{dE_{\bar{\nu}}}\Big|}\Bigg|_{E_{\bar{\nu}}
  = E_r},
\end{eqnarray}
where $E_r$ is the root of the equation $f(E_r) = 0$. It is equal to
\begin{eqnarray}\label{eq:A.18}
\hspace{-0.3in}E_r = \frac{E_0 - E_e}{\displaystyle 1 -
  \frac{1}{m_n}(E_e - k_e\cos\vartheta_{e\bar{\nu}})}.
\end{eqnarray}
Then, we get \cite{Ivanov2013}
\begin{eqnarray}\label{eq:lA.19}
\hspace{-0.3in}&&\frac{m_n}{E_p}\,\frac{1}{\displaystyle
  \Big|\frac{df(E_{\bar{\nu}})}{dE_{\bar{\nu}}}\Big|}\Bigg|_{E_{\bar{\nu}}
  = E_r} = \frac{1}{\displaystyle 1 - \frac{1}{m_n}(E_e -
  k_e\cos\vartheta_{e\bar{\nu}})}.
\end{eqnarray}
The exact expression for the function $\Phi_n(\vec{k}_e,
\cos \theta_{e\bar{\nu}})$ is 
\begin{eqnarray}\label{eq:A.20}
\hspace{-0.3in}\Phi_n(k_e, \cos \theta_{e\bar{\nu}}) =
\frac{1}{\displaystyle \Big(1 - \frac{1}{m_n}(E_e -
  k_e\cos\vartheta_{e\bar{\nu}})\Big)^3}.
\end{eqnarray}
Replacing $\Phi_n(k_e, \cos \theta_{e\bar{\nu}})$ by
$\Phi_n(\vec{k}_e, \vec{k}_{\bar{\nu}})$ \cite{Ivanov2013},
expanding the r.h.s. of Eq.(\ref{eq:A.20}) in powers of $1/m_N$ and
keeping the contributions of order $1/m^2_N$ we get
\begin{eqnarray}\label{eq:A.21}
\hspace{-0.3in} \Phi_n(\vec{k}_e, \vec{k}_{\bar{\nu}}) = 1 +
3\, \frac{E_e}{m_N}\,\Big(1 - \frac{\vec{k}_e \cdot
  \vec{k}_{\bar{\nu}}}{E_e E_{\bar{\nu}}}\Big) + 6\,
\frac{E^2_e}{m^2_N}\Big(1 - \frac{\vec{k}_e \cdot
  \vec{k}_{\bar{\nu}}}{E_e E_{\bar{\nu}}}\Big)\Big(1 - \frac{\vec{k}_e
  \cdot \vec{k}_{\bar{\nu}}}{E_e E_{\bar{\nu}}} - \frac{1}{4}\,
\frac{E_0}{E_e}\Big),
\end{eqnarray}
where we have denoted $|\vec{k}_{\bar{\nu}}| = E_{\bar{\nu}} = E_0 -
E_e$ \cite{Ivanov2013}.

Thus, the corrections of order $O(E^2_e/m^2_N)$, caused by weak
magnetism and proton recoil, appear in the neutron lifetime and the
correlation coefficients of the neutron beta decay from i) the
corrections of order $O(E^2_0/m^2_N)$ to the amplitude of the neutron
beta decay, given by Eq.(\ref{eq:A.15}), ii) the corrections of order
$O(E^2_e/m^2_N)$ to the phase--volume of the neutron beta decay, iii)
the corrections of order $O(E^2_e/m^2_N)$, caused by the quadratic and
crossing terms of order $O(E_e/m_N)$ in Eq.(\ref{eq:A.14}) without
$\delta {\cal M}_n$, iv) the corrections of order $O(E^2_e/m^2_N)$,
obtained by the multiplication of the corrections of order
$O(E_e/m_N)$ to the electron--energy and angular distribution,
calculated without account for the contributions of the phase--volume
of the neutron beta decay, by the second term of order $O(E_e/m_N)$ in
Eq.(\ref{eq:A.21}) appearing from the phase--volume of the neutron
beta decay, and v) the corrections, induced by the expansion of
$1/\zeta(E_e)$ in powers of $1/m_N$ and $1/m^2_N$, respectively.

Of course, having considered corrections of order $O(E^2_e/m^2_N) \sim
10^{-5}$ we may, in principle, take into account the contributions of
the isovector and axial--vector nucleon form factors
\cite{Bernard1995, Liesenfeld1999, Leitner2006, Ivanov2018x,
  Ivanov2019b}. Taking the contributions of these form factors in the
dipole approximation we rewrite Eq.(\ref{eq:A.15}) as follows
\begin{eqnarray}\label{eq:A.22}
  \hspace{-0.3in}&&\delta {\cal M}_n = 2 m_n \,\Big\{ - 2\,\Big(
  \frac{q^2}{M^2_V}\, [\varphi^{\dagger}_p \varphi_n][\bar{u}_e
    \gamma^0 (1 - \gamma^5) v_{\bar{\nu}}] + g_A\, \frac{q^2}{M^2_A}\,
       [\varphi^{\dagger}_p \vec{\sigma}\, \varphi_n] \cdot [\bar{u}_e
         \vec{\gamma}\, (1 - \gamma^5) v_{\bar{\nu}}]\Big) +
       \frac{E^2_0}{8 m^2_N} \Big[(2 \kappa + 1)\, \frac{q^2}{E^2_0}
         \nonumber\\
 \hspace{-0.3in}&& \times \,[\varphi^{\dagger}_p \varphi_n][\bar{u}_e
   \gamma^0 (1 - \gamma^5) v_{\bar{\nu}}] - g_A \frac{q^2 -
   2m^2_e}{E^2_0} \, [\varphi^{\dagger}_p\vec{\sigma}\,\varphi_n]\cdot
        [\bar{u}_e \vec{\gamma}\, (1 - \gamma^5) v_{\bar{\nu}}] - 2
        \kappa\, \frac{m_e}{E_0} \,[\varphi^{\dagger}_p\varphi_n]
                 [\bar{u}_e (1 - \gamma^5)v_{\bar{\nu}}]\Big]\Big\},
\end{eqnarray}
where in front of the brackets in the first term on the first line the
factor 2 comes from the dipole approximation of the isovector and
axial--vector form factors.  The slope-parameters $M_V$ and $M_A$ we
relate to the charge radius of the proton $r_p = 0.841\,{\rm fm}$
\cite{Pohl2010, Antognini2013} and the axial radius $r_A = 0.635\,{\rm
  fm}$ of the nucleon \cite{Liesenfeld1999, Ivanov2019c},
respectively. This gives $M_V = \sqrt{12}/r_p = 813\,{\rm MeV}$ and
$M_A = \sqrt{12}/r_A = 1077\, {\rm MeV}$. As a result, the
electron--energy and angular distribution Eq.(\ref{eq:2}) acquires the
following correction
\begin{eqnarray}\label{eq:A.23}
\hspace{-0.15in}&&\frac{d^5 \delta \lambda_n(E_e, \vec{k}_e,
  \vec{k}_{\bar{\nu}}, \vec{\xi}_n, \vec{\xi}_e)}{dE_e d\Omega_e
  d\Omega_{\bar{\nu}}} \propto \bigg\{ \Big[- \frac{8}{1 + 3
    g^2_A}\,\Big(\frac{E^2_0}{M^2_V} + 3 g^2_A
  \frac{E^2_0}{M^2_A}\Big) - \frac{8 g_A}{1 + 3
    g^2_A}\,\Big(\frac{E^2_0}{M^2_V} + (1 + 2 g_A)\,
  \frac{E^2_0}{M^2_A}\Big)\, \frac{\vec{\xi}_n \cdot
    \vec{k}_{\bar{\nu}}}{E_{\bar{\nu}}}\Big]\Big(1 - \frac{\vec{\xi}_e
  \cdot \vec{k}_e}{E_e}\Big) \nonumber\\
 \hspace{-0.3in}&& + \Big[- \frac{8 g_A}{1 + 3
     g^2_A}\,\Big(\frac{E^2_0}{M^2_V} + (1 - 2 g_A)\,
   \frac{E^2_0}{M^2_A}\Big)\, \vec{\xi}_n - \frac{8}{1 + 3
     g^2_A}\,\Big(\frac{E^2_0}{M^2_V} - g^2_A \,
   \frac{E^2_0}{M^2_A}\Big)\,
   \frac{\vec{k}_{\bar{\nu}}}{E_{\bar{\nu}}} \Big] \cdot
 \Big(\frac{\vec{k}_e}{E_e} - \frac{m_e}{E_e}\, \vec{\xi}_e -
 \frac{\vec{k}_e (\vec{\xi}_e \cdot \vec{k}_e)}{(E_e + m_e)
   E_e}\Big)\bigg\}\nonumber\\
 \hspace{-0.3in}&& \times \,\frac{E_e}{E_0}\,\Big(1 -
 \frac{E_e}{E_0}\Big)\,\Big(1 - \frac{\vec{k}_e \cdot
   \vec{k}_{\bar{\nu}}}{E_e E_{\bar{\nu}}}\Big).
\end{eqnarray}
Using the numerical values of the parameters $g_A = 1.2764$
\cite{Abele2018}, $M_V = 813\,{\rm MeV}$, $M_A = 1077\, {\rm MeV}$ and
$E_0 = 1.2926\, {\rm MeV}$ we get
\begin{eqnarray}\label{eq:A.24}
\hspace{-0.15in}&&\frac{d^5 \delta \lambda_n(E_e, \vec{k}_e,
  \vec{k}_{\bar{\nu}}, \vec{\xi}_n, \vec{\xi}_e)}{dE_e d\Omega_e
  d\Omega_{\bar{\nu}}} \propto \bigg\{ \Big[- 1.30 \times 10^{-5} -
  1.33 \times 10^{-5}\, \frac{\vec{\xi}_n \cdot
    \vec{k}_{\bar{\nu}}}{E_{\bar{\nu}}}\Big]\Big(1 - \frac{\vec{\xi}_e
  \cdot \vec{k}_e}{E_e}\Big) \nonumber\\
 \hspace{-0.3in}&& + \Big[- 5.05 \times 10^{-7}\, \vec{\xi}_n - 2.46
   \times 10^{-7} \, \frac{\vec{k}_{\bar{\nu}}}{E_{\bar{\nu}}} \Big]
 \cdot \Big(\frac{\vec{k}_e}{E_e} - \frac{m_e}{E_e}\, \vec{\xi}_e -
 \frac{\vec{k}_e (\vec{\xi}_e \cdot \vec{k}_e)}{(E_e + m_e)
   E_e}\Big)\bigg\}\nonumber\\
 \hspace{-0.3in}&& \times \,\frac{E_e}{E_0}\,\Big(1 -
 \frac{E_e}{E_0}\Big)\,\Big(1 - \frac{\vec{k}_e \cdot
   \vec{k}_{\bar{\nu}}}{E_e E_{\bar{\nu}}}\Big).
\end{eqnarray}
Having neglected the contributions of the terms of order of a few
parts of $10^{-7}$ we obtain
\begin{eqnarray}\label{eq:A.25}
\hspace{-0.15in}&&\frac{d^5 \delta \lambda_n(E_e, \vec{k}_e,
  \vec{k}_{\bar{\nu}}, \vec{\xi}_n, \vec{\xi}_e)}{dE_e d\Omega_e
  d\Omega_{\bar{\nu}}} \propto \bigg\{ \Big[- \frac{8}{1 + 3
    g^2_A}\,\Big(\frac{E^2_0}{M^2_V} + 3 g^2_A
  \frac{E^2_0}{M^2_A}\Big) + \frac{8}{1 + 3
    g^2_A}\,\Big(\frac{E^2_0}{M^2_V} + 3 g^2_A
  \frac{E^2_0}{M^2_A}\Big)\,\frac{\vec{k}_e \cdot
    \vec{k}_{\bar{\nu}}}{E_e E_{\bar{\nu}}} - \frac{8 g_A}{1 + 3
   g^2_A}\nonumber\\
 \hspace{-0.3in}&& \times \,\Big(\frac{E^2_0}{M^2_V} + (1 + 2 g_A)\,
 \frac{E^2_0}{M^2_A}\Big)\, \frac{\vec{\xi}_n \cdot
   \vec{k}_{\bar{\nu}}}{E_{\bar{\nu}}} + \frac{8}{1 + 3
   g^2_A}\,\Big(\frac{E^2_0}{M^2_V} + 3 g^2_A
 \frac{E^2_0}{M^2_A}\Big)\, \frac{\vec{\xi}_e \cdot \vec{k}_e}{E_e} -
 \frac{8}{1 + 3 g^2_A}\,\Big(\frac{E^2_0}{M^2_V} + 3 g^2_A
 \frac{E^2_0}{M^2_A}\Big)\, \Big(1 + \frac{m_e}{E_e}\Big) \nonumber\\
 \hspace{-0.3in}&& \times \,\frac{(\vec{\xi}_e \cdot
   \vec{k}_e)(\vec{k}_e \cdot \vec{k}_{\bar{\nu}})}{(E_e + m_e) E_e
   E_{\bar{\nu}}}+ \frac{8 g_A}{1 + 3 g^2_A}\,\Big(\frac{E^2_0}{M^2_V}
 + (1 + 2 g_A)\, \frac{E^2_0}{M^2_A}\Big)\, \frac{(\vec{\xi}_n \cdot
   \vec{k}_{\bar{\nu}})(\vec{\xi}_e \cdot \vec{k}_e)}{E_e
   E_{\bar{\nu}}} + \frac{8 g_A}{1 + 3
   g^2_A}\,\Big(\frac{E^2_0}{M^2_V} + (1 + 2 g_A)\,
 \frac{E^2_0}{M^2_A}\Big)\nonumber\\
 \hspace{-0.3in}&& \times \Big(\frac{(\vec{\xi}_n \cdot
   \vec{k}_{\bar{\nu}})(\vec{k}_e \cdot \vec{k}_{\bar{\nu}})}{E_e
   E^2_{\bar{\nu}}} - \frac{1}{3} \frac{\vec{\xi}_n \cdot
   \vec{k}_e}{E_e}\Big) - \frac{8 g_A}{1 + 3 g^2_A}
 \Big(\frac{E^2_0}{M^2_V} + (1 + 2 g_A) \frac{E^2_0}{M^2_A}\Big) \Big(
 \frac{(\vec{\xi}_n \cdot \vec{k}_{\bar{\nu}})(\vec{\xi}_e \cdot
   \vec{k}_e)(\vec{k}_e \cdot \vec{k}_{\bar{\nu}})}{E^2_e
   E^2_{\bar{\nu}}} - \frac{1}{3} \frac{(\vec{\xi}_n \cdot
   \vec{k}_e)(\vec{\xi}_e \cdot
   \vec{k}_e)}{E^2_e}\Big)\bigg\}\nonumber\\
 \hspace{-0.3in}&& \times \,\frac{E_e}{E_0}\,\Big(1 -
 \frac{E_e}{E_0}\Big).
\end{eqnarray}
The contributions to the correlation function $\zeta(E_e)$ and the
correlation coefficients we define as $\zeta(E_e)_{\rm FF}$,
$a(E_e)_{\rm FF}$, $B(E_e)_{\rm FF}$, $G(E_e)_{\rm FF}$ and
$K_e(E_e)_{\rm FF}$, respectively.

\vspace{-0.1in}
\section*{Appendix B: Electron--energy and angular distributions of
  the neutron beta decay, beyond the structure introduced by Jackson
  {\it et al.} \cite{Jackson1957} }
\renewcommand{\theequation}{B-\arabic{equation}}
\setcounter{equation}{0}
\vspace{-0.1in}
In this Appendix we adduce the contributions of corrections, caused by
weak magnetism and proton recoil to next-to-leading and to
next-to-next-to-leading order in the large nucleon mass $m_N$
expansion, which go beyond the correlation structure of the
electron--energy and angular distribution of the neutron beta decay,
introduced by Jackson {\it et al.} \cite{Jackson1957}. We give
\begin{eqnarray}\label{eq:B.1}
\hspace{-0.15in}&&\frac{d^5 \lambda_n(E_e, \vec{k}_e,
  \vec{k}_{\bar{\nu}}, \vec{\xi}_n, \vec{\xi}_e)}{dE_e d\Omega_e
  d\Omega_{\bar{\nu}}}\Big|_{(\rm NLO)} = (1 + 3
g^2_A)\,\frac{G^2_F|V_{ud}|^2}{32\pi^5}\,(E_0 - E_e)^2 \,\sqrt{E^2_e -
  m^2_e}\, E_e\,F(E_e, Z = 1)\,\zeta(E_e) \nonumber\\
\hspace{-0.15in}&& \times \,\frac{E_e}{m_N}\,\Big\{ -
3\, \frac{1 - g^2_A}{1 + 3 g^2_A}
\,\Big(\frac{(\vec{k}_e\cdot \vec{k}_{\bar{\nu}})^2}{E^2_e
  E^2_{\bar{\nu}}} - \frac{1}{3}\,\frac{k^2_e}{E^2_e}\Big) +
3\,\frac{1 - g^2_A}{1 + 3 g^2_A}\, \Big(\frac{(\vec{\xi}_e\cdot
  \vec{k}_{\bar{\nu}})(\vec{k}_e\cdot \vec{k}_{\nu})}{E_e
  E^2_{\bar{\nu}}} - \frac{1}{3}\,\frac{\vec{\xi}_e\cdot
  \vec{k}_e}{E_e}\,\Big)\,\frac{m_e}{E_e} \nonumber\\
\hspace{-0.15in}&& + 3\,\frac{1 - g^2_A}{1 + 3 g^2_A}\,
\Big(\frac{(\vec{\xi}_e\cdot \vec{k}_e)(\vec{k}_e\cdot
  \vec{k}_{\bar{\nu}})^2}{(E_e + m_e)E^2_e E^2_{\bar{\nu}}} -
\frac{1}{3}\,\Big(1 - \frac{m_e}{E_e}\Big)\,\frac{\vec{\xi}_e\cdot
  \vec{k}_e}{E_e}\,\Big)\Big\}
\end{eqnarray}
and 
\begin{eqnarray*}
\hspace{-0.15in}&&\frac{d^5 \lambda^{(1)}_n(E_e, \vec{k}_e,
  \vec{k}_{\bar{\nu}}, \vec{\xi}_n, \vec{\xi}_e)}{dE_e d\Omega_e
  d\Omega_{\bar{\nu}}}\Big|_{(\rm N^2LO)} = (1 + 3
g^2_A)\,\frac{G^2_F|V_{ud}|^2}{32\pi^5}\,(E_0 - E_e)^2 \,\sqrt{E^2_e -
  m^2_e}\, E_e\,F(E_e, Z = 1)\,\zeta(E_e) \nonumber\\
\hspace{-0.15in}&& \times \,\bigg\{\frac{E^2_0}{2 m^2_N}\, \Big\{ -
\frac{1}{1 + 3 g^2_A}\, (g_A + 2\kappa + 1)\, \frac{E_e
  E_{\bar{\nu}}}{E^2_0}\Big(\frac{(\vec{k}_e\cdot
  \vec{k}_{\bar{\nu}})^2}{E^2_e E^2_{\bar{\nu}}} - \frac{1}{3}\,
\frac{k^2_e}{E^2_e}\Big) - \frac{1}{1 + 3 g^2_A}\, (g^2_A + 2 \kappa +
1) \, \frac{E_{\bar{\nu}}}{E_e + m_e}\nonumber\\
\hspace{-0.3in}&&\times \,\Big(
\frac{(\vec{k}_e\cdot \vec{k}_{\bar{\nu}})^2}{E^2_e E^2_{\bar{\nu}}} -
\frac{1}{3}\, \frac{k^2_e}{E^2_e}\Big)\, \frac{\vec{\xi}_e \cdot
  \vec{k}_e}{E_e} - \frac{1}{1 + 3 g^2_A}\, (g^2_A + 2\kappa + 1) \frac{m_e
  E_{\bar{\nu}}}{E^2_0}\Big(\frac{(\vec{\xi}_e \cdot
  \vec{k}_{\bar{\nu}})(\vec{k}_e \cdot \vec{k}_{\bar{\nu}})}{E_e
  E^2_{\bar{\nu}}} - \frac{1}{3}\,\frac{\vec{\xi}_e \cdot
  \vec{k}_e}{E_e} \Big) \nonumber\\
\hspace{-0.15in}&& + \frac{1}{1 + 3 g^2_A}\, \Big( - 2 g_A(g_A +
\kappa)\,\frac{E_e E_{\bar{\nu}}}{E^2_0} + g_A (g_A - \kappa - 1)\,
\frac{m^2_e}{E^2_0}\Big) \, \frac{(\vec{\xi}_n \cdot
  \vec{k}_{\bar{\nu}})(\vec{\xi}_e \cdot \vec{k}_e)}{E_e
  E_{\bar{\nu}}}+ \frac{1}{1 + 3 g^2_A} \, 2 \kappa g_A\, \frac{E_e
  E_{\bar{\nu}}}{E^2_0}\nonumber\\
\hspace{-0.15in}&& \times \,\Big( \frac{(\vec{\xi}_n \cdot
  \vec{k}_{\bar{\nu}})(\vec{k}_e \cdot \vec{\xi}_e)(\vec{k}_e \cdot
  \vec{k}_{\bar{\nu}})}{E^2_e E^2_{\bar{\nu}}} - \frac{1}{3}\,
\frac{(\vec{\xi}_n \cdot \vec{k}_e)(\vec{k}_e \cdot \vec{\xi}_e)}{
  E^2_e } \Big) + \frac{1}{1 + 3 g^2_A}\, \frac{m_e}{E_0}\Big(2 g_A
(g_A + \kappa)\, \frac{ E_{\bar{\nu}}}{E_0} - \kappa g_A\Big)\,
\frac{(\vec{\xi}_n \cdot \vec{\xi}_e)(\vec{k}_e \cdot
  \vec{k}_{\bar{\nu}})}{E_e E_{\bar{\nu}}} \nonumber\\
\end{eqnarray*}
\begin{eqnarray}\label{eq:B.2}
\hspace{-0.15in}&& + \frac{1}{1 + 3 g^2_A}\, \Big( \kappa\, g_A
\frac{m_e}{E_0}\Big) \, \frac{(\vec{\xi}_n \cdot
  \vec{k}_e)(\vec{\xi}_e \cdot \vec{k}_{\bar{\nu}})}{E_e
  E_{\bar{\nu}}} + \frac{1}{1 + 3 g^2_A} \, \Big(2 \kappa g_A \,
\frac{E_e E_{\bar{\nu}}}{E^2_0}\Big) \, \frac{(\vec{\xi}_n \cdot
  \vec{k}_e)(\vec{k}_e \cdot \vec{\xi}_e)(\vec{k}_e \cdot
  \vec{k}_{\bar{\nu}})}{(E_e + m_e) E^2_e E_{\bar{\nu}}}\Big\}
+ \Big\{\frac{8 g_A}{1 + 3 g^2_A} \nonumber\\
\hspace{-0.15in}&& \times \,\Big(\frac{E^2_0}{M^2_V} +
  (1 + 2 g_A)\, \frac{E^2_0}{M^2_A}\Big)\, \frac{(\vec{\xi}_n \cdot
    \vec{k}_{\bar{\nu}})(\vec{\xi}_e \cdot \vec{k}_e)}{E_e
    E_{\bar{\nu}}} + \frac{8 g_A}{1 + 3
    g^2_A}\,\Big(\frac{E^2_0}{M^2_V} + (1 + 2 g_A)\,
  \frac{E^2_0}{M^2_A}\Big) \Big(\frac{(\vec{\xi}_n \cdot
    \vec{k}_{\bar{\nu}})(\vec{k}_e \cdot \vec{k}_{\bar{\nu}})}{E_e
    E^2_{\bar{\nu}}} - \frac{1}{3} \frac{\vec{\xi}_n \cdot
    \vec{k}_e}{E_e}\Big) \nonumber\\
\hspace{-0.15in}&&  - \frac{8 g_A}{1 + 3 g^2_A}
  \Big(\frac{E^2_0}{M^2_V} + (1 + 2 g_A) \frac{E^2_0}{M^2_A}\Big)
  \Big( \frac{(\vec{\xi}_n \cdot \vec{k}_{\bar{\nu}})(\vec{\xi}_e
    \cdot \vec{k}_e)(\vec{k}_e \cdot \vec{k}_{\bar{\nu}})}{E^2_e
    E^2_{\bar{\nu}}} - \frac{1}{3} \frac{(\vec{\xi}_n \cdot
    \vec{k}_e)(\vec{\xi}_e \cdot
    \vec{k}_e)}{E^2_e}\Big)\Big\}\,\frac{E_e}{E_0}\,\Big(1 -
  \frac{E_e}{E_0}\Big)\bigg\}
\end{eqnarray}
and
\begin{eqnarray}\label{eq:B.3}
\hspace{-0.15in}&&\frac{d^5 \lambda^{(2)}_n(E_e, \vec{k}_e,
  \vec{k}_{\bar{\nu}}, \vec{\xi}_n, \vec{\xi}_e)}{dE_e d\Omega_e
  d\Omega_{\bar{\nu}}}\Big|_{(\rm N^2LO)} = (1 + 3
g^2_A)\,\frac{G^2_F|V_{ud}|^2}{32\pi^5}\,(E_0 - E_e)^2 \,\sqrt{E^2_e -
  m^2_e}\, E_e\,F(E_e, Z = 1)\,\zeta(E_e) \nonumber\\
\hspace{-0.15in}&& \times \, 6\, \frac{E^2_e}{m^2_N} \Big\{ \Big[1 -
  2\, \frac{1 - g^2_A}{1 + 3 g^2_A}\, \Big(1 - \frac{1}{8}\,
  \frac{E_0}{E_e}\Big)\Big] \,\Big(\frac{(\vec{k}_e\cdot
  \vec{k}_{\bar{\nu}})^2}{E^2_e E^2_{\bar{\nu}}} - \frac{1}{3}\,
\frac{k^2_e}{E^2_e}\Big) + 2 \, \frac{g_A(1 - g_A)}{1 + 3 g^2_A}
\Big(\frac{(\vec{k}_e\cdot \vec{k}_{\bar{\nu}})^2}{E^2_e
  E^2_{\bar{\nu}}} - \frac{1}{3}\, \frac{k^2_e}{E^2_e}\Big)
\frac{\vec{\xi}_n \cdot \vec{k}_e}{E_e} \nonumber\\
\hspace{-0.15in}&& + \Big[ - 1 + 2 \, \frac{1 - g^2_A}{1 + 3
    g^2_A}\,\Big(1 - \frac{1}{8}\, \frac{E_0}{E_e}\Big)\,
  \frac{E_e}{E_e + m_e}\Big] \, \Big(\frac{(\vec{k}_e\cdot
  \vec{k}_{\bar{\nu}})^2}{E^2_e E^2_{\bar{\nu}}} - \frac{1}{3}\,
\frac{k^2_e}{E^2_e}\Big)\, \frac{\vec{\xi}_e \cdot \vec{k}_e}{E_e} +
2\, \frac{1 - g^2_A}{1 + 3 g^2_A}\,\frac{m_e}{E_e}\,\Big(1 -
\frac{1}{8}\, \frac{E_0}{E_e}\Big) \nonumber\\
\hspace{-0.15in}&& \times \Big(\frac{(\vec{\xi}_e \cdot
  \vec{k}_{\bar{\nu}})(\vec{k}_e \cdot \vec{k}_{\bar{\nu}})}{E_e
  E^2_{\bar{\nu}}}- \frac{1}{3}\, \frac{\vec{\xi}_e \cdot
  \vec{k}_e}{E_e}\Big) - 2 \, \frac{g_A(1 - g_A)}{1 + 3 g^2_A}\,
\frac{m_e}{E_e}\,\Big(\frac{(\vec{k}_e\cdot
  \vec{k}_{\bar{\nu}})^2}{E^2_e E^2_{\bar{\nu}}} - \frac{1}{3}\,
\frac{k^2_e}{E^2_e}\Big) \, (\vec{\xi}_n \cdot \vec{\xi}_e) + 4\,
\frac{g_A(1 + g_A)}{1 + 3 g^2_A}\, \Big(1 - \frac{1}{8}\,
\frac{E_0}{E_e}\Big)\nonumber\\
\hspace{-0.15in}&& \times \Big(\frac{(\vec{\xi}_n \cdot
  \vec{k}_{\bar{\nu}})(\vec{k}_e \cdot \vec{\xi}_e)(\vec{k}_e \cdot
  \vec{k}_{\bar{\nu}})}{E^2_e E^2_{\bar{\nu}}} - \frac{1}{3}\,
\frac{(\vec{\xi}_n \cdot \vec{k}_e)(\vec{k}_e \cdot
  \vec{\xi}_e)}{E^2_e}\Big) + 2 \, \frac{g_A(1 - g_A)}{1 + 3 g^2_A}\,
\frac{m_e}{E_e}\,\Big(\frac{(\vec{k}_e\cdot
  \vec{k}_{\bar{\nu}})^2}{E^2_e E^2_{\bar{\nu}}} - \frac{1}{3}\,
\frac{k^2_e}{E^2_e}\Big) \,\frac{(\vec{\xi}_n \cdot
  \vec{k}_e)(\vec{k}_e \cdot \vec{\xi}_e)}{(E_e + m_e) E_e}
\nonumber\\
\hspace{-0.15in}&& + 4\, \frac{g_A(1 - g_A)}{1 + 3 g^2_A}\,
\frac{m_e}{E_e}\, \Big(1 - \frac{1}{8}\, \frac{E_0}{E_e}\Big)\,
\frac{(\vec{\xi}_n \cdot \vec{\xi}_e)(\vec{k}_e \cdot
  \vec{k}_{\bar{\nu}})}{E_e E_{\bar{\nu}}} - 2 \, \frac{g_A(1 +
  g_A)}{1 + 3 g^2_A} \Big(1 - \frac{1}{4}\, \frac{E_0}{E_e}\Big)
\frac{(\vec{\xi}_n \cdot \vec{k}_{\bar{\nu}})(\vec{k}_e \cdot
  \vec{\xi}_e)}{E_e E_{\bar{\nu}}}\nonumber\\
\hspace{-0.15in}&& + 4\, \frac{g_A (1 - g_A)}{1 + 3 g^2_A}\, \Big(1 -
\frac{1}{8}\, \frac{E_0}{E_e}\Big)\, \frac{(\vec{\xi}_n \cdot
  \vec{k}_e) (\vec{k}_e \cdot \vec{\xi}_e) (\vec{k}_e \cdot
  \vec{k}_{\bar{\nu}})}{(E_e + m_e) E^2_e E_{\bar{\nu}}} + 2 \,
\frac{g_A(1 + g_A)}{1 + 3 g^2_A}\frac{(\vec{\xi}_n \cdot
  \vec{k}_{\bar{\nu}})(\vec{k}_e \cdot \vec{k}_{\bar{\nu}})^2}{E^2_e
  E^3_{\bar{\nu}}}- 2 \, \frac{g_A(1 + g_A)}{1 + 3 g^2_A}\nonumber\\
\hspace{-0.15in}&& \times \frac{(\vec{\xi}_n \cdot
  \vec{k}_{\bar{\nu}})(\vec{k}_e \cdot \vec{\xi}_e)(\vec{k}_e \cdot
  \vec{k}_{\bar{\nu}})^2}{E^3_e E^3_{\bar{\nu}}} - \frac{1 - g^2_A}{1
  + 3 g^2_A}\, \frac{m_e}{E_e}\, \frac{(\vec{\xi}_e \cdot
  \vec{k}_{\bar{\nu}})(\vec{k}_e \cdot \vec{k}_{\bar{\nu}})^2}{E^2_e
  E^3_{\bar{\nu}}} - \frac{1 - g^2_A}{1 + 3 g^2_A}\,
\frac{(\vec{\xi}_e \cdot \vec{k}_e)(\vec{k}_e \cdot
  \vec{k}_{\bar{\nu}})^3}{(E_e + m_e) E^3_e E^3_{\bar{\nu}}}- \frac{1
  - g^2_A}{1 + 3 g^2_A}\, \frac{(\vec{k}_e \cdot
  \vec{k}_{\bar{\nu}})^3}{ E^3_e E^3_{\bar{\nu}}}\Big\}\nonumber\\
\hspace{-0.15in}&& 
\end{eqnarray}
and
\begin{eqnarray*}
\hspace{-0.30in}&&\frac{d^5 \lambda^{(3)}_n(E_e, \vec{k}_e,
  \vec{k}_{\bar{\nu}}, \vec{\xi}_n, \vec{\xi}_e)}{dE_e d\Omega_e
  d\Omega_{\bar{\nu}}}\Big|_{(\rm N^2LO)} = (1 + 3
g^2_A)\,\frac{G^2_F|V_{ud}|^2}{32\pi^5}\,(E_0 - E_e)^2 \,\sqrt{E^2_e -
  m^2_e}\, E_e\,F(E_e, Z = 1)\,\zeta(E_e) \nonumber\\
\hspace{-0.30in}&& \times \,\frac{E^2_0}{2 m^2_N}\,\Big\{ - \frac{1}{1
  + 3 g^2_A}\, \Big[(g^2_A + (\kappa + 1)^2)\,
  \frac{E_{\bar{\nu}}}{E_0}\Big] \Big(\frac{(\vec{k}_e \cdot
  \vec{k}_{\bar{\nu}})^2}{E^2_e E^2_{\bar{\nu}}} -
\frac{1}{3}\,\frac{k^2_e}{E^2_e}\Big) + \frac{1}{1 + 3 g^2_A}\,
 \Big[(g^2_A + (\kappa + 1)^2)\,
  \frac{m_e}{E_0}\, \frac{E_{\bar{\nu}}}{E_0}\Big]\nonumber\\
\hspace{-0.30in}&& \times \Big(\frac{(\vec{\xi}_e \cdot
  \vec{k}_{\bar{\nu}})(\vec{k}_e \cdot \vec{k}_{\bar{\nu}})}{E_e
  E^2_{\bar{\nu}}} - \frac{1}{3}\, \frac{\vec{\xi}_e \cdot
  \vec{k}_e}{E_e}\Big) + \frac{1}{1 + 3 g^2_A}\,\Big[(g^2_A + (\kappa
  + 1)^2)\, \frac{E_e E_{\bar{\nu}}}{E^2_0}\Big] \Big(\frac{(\vec{k}_e
  \cdot \vec{k}_{\bar{\nu}})^2}{E^2_e E^2_{\bar{\nu}}} -
\frac{1}{3}\,\frac{k^2_e}{E^2_e}\Big)\, \frac{\vec{\xi}_e \cdot
  \vec{k}_e}{E_e + m_e} \nonumber\\
\hspace{-0.30in}&& + \frac{1}{1 + 3 g^2_A}\, \,
\frac{m_e}{E_e}\, \Big[(\kappa + 1)\big(g_A + (\kappa + 1)\big)\,
\frac{E^2_{\bar{\nu}}}{E^2_0} + \big(g_A + (\kappa + 1)\big)\,
\frac{E_e E_{\bar{\nu}}}{E^2_0} + g_A \big(g_A + (\kappa +
1)\big)\,\frac{E_{\bar{\nu}}}{E_0}\Big] \nonumber\\
\hspace{-0.30in}&& \times \Big(\frac{(\vec{\xi}_n \cdot
  \vec{k}_{\bar{\nu}})(\vec{k}_{\bar{\nu}} \cdot
  \vec{\xi}_e)}{E^2_{\bar{\nu}}} - \frac{1}{3}\, \vec{\xi}_n \cdot
\vec{\xi}_e\Big) + \frac{1}{1 + 3 g^2_A}\, \Big[g_A \big(g_A + (\kappa
  + 1)\big)\, \frac{E_{\bar{\nu}}}{E_0} + \big(g_A + (\kappa +
  1)\big)\, \frac{E_e E_{\bar{\nu}}}{E^2_0} \nonumber\\
\hspace{-0.30in}&& + (\kappa + 1)\big(g_A + (\kappa +
1)\big)\,\frac{E^2_{\bar{\nu}}}{E^2_0} - (\kappa + 2)\big(g_A +
(\kappa + 1)\big)\,\frac{E_e E_{\bar{\nu}}}{E^2_0}\Big(1 +
\frac{m_e}{E_e}\Big)\Big] \Big(\frac{(\vec{\xi}_n \cdot
  \vec{k}_{\bar{\nu}})(\vec{k}_e \cdot \vec{k}_{\bar{\nu}})}{E_e
  E^2_{\bar{\nu}}} - \frac{1}{3}\, \frac{\vec{\xi}_n \cdot
  \vec{k}_e}{E_e}\Big)\nonumber\\
\hspace{-0.30in}&& \times \frac{\vec{k}_e \cdot \vec{\xi}_e}{E_e +
  m_e} + \frac{1}{1 + 3 g^2_A}\, \Big[\Big(- g^2_A + (g_A + 1)(\kappa
  + 1)\, \frac{m^2_e}{E^2_0}\Big) - 2 (\kappa + 1)(g_A + 1)\,
  \frac{E^2_e}{E^2_0} + g_A\big(g_A + (\kappa + 1)\big)\,
  \frac{E_e}{E_0} \nonumber\\
\hspace{-0.30in}&& - 4 g_A (\kappa + 1)\, \frac{E_{\bar{\nu}}}{E_0} +
(\kappa + 1)^2\, \frac{E_e E_{\bar{\nu}}}{E^2_0} - (\kappa + 1)^2\,
\frac{E^2_{\bar{\nu}}}{E^2_0}\Big]\, \frac{(\vec{\xi}_n \cdot
  \vec{k}_{\bar{\nu}})(\vec{k}_e \cdot \vec{\xi}_e)}{E_e
  E_{\bar{\nu}}} + \frac{1}{1 + 3 g^2_A}\, 
\frac{m_e}{E_e}\Big[- g_A \big(g_A - (\kappa + 2)\big)\,
  \frac{E_e}{E_0} \nonumber\\
\hspace{-0.30in}&& - (\kappa + 1)\, \frac{E^2_e}{E^2_0} - (\kappa + 1)
(2 g_A - 1)\, \frac{E_e E_{\bar{\nu}}}{E^2_0}\Big]\,
\frac{(\vec{\xi}_n \cdot \vec{\xi}_e)(\vec{k}_e \cdot
  \vec{k}_{\bar{\nu}})}{E_e E_{\bar{\nu}}} + \frac{1}{1 + 3 g^2_A}\,
\, \frac{m_e}{E_e}\Big[(\kappa + 1)(g_A + \kappa)\, \frac{E_e
    E_{\bar{\nu}}}{E^2_0} + g_A\,\frac{E^2_e}{E^2_0}\Big]\nonumber\\
\end{eqnarray*}
\begin{eqnarray}\label{eq:B.4}  
\hspace{-0.30in}&& \times \,\frac{(\vec{\xi}_n \cdot
  \vec{k}_e)(\vec{k}_{\bar{\nu}}\cdot \vec{\xi}_e)}{E_e E_{\bar{\nu}}}
+ \frac{1}{1 + 3 g^2_A}\, \, \Big[\kappa \big(g_A - (\kappa +
  1)\big)\, \frac{m_e}{E_0}\, \frac{E_e}{E_0} - g_A \big(g_A - (\kappa
  + 1)\big) \, \frac{E_e}{E_0} + (\kappa + 1) \big(g_A - (\kappa +
  1)\big) \, \frac{E^2_e}{E^2_0} \nonumber\\ 
\hspace{-0.30in}&& + (\kappa + 1)\big(3 g_A + (\kappa + 1)\big)\,
\frac{E_e E_{\bar{\nu}}}{E^2_0}\Big]\, \frac{(\vec{\xi}_n \cdot
  \vec{k}_e)(\vec{k}_e \cdot \vec{\xi}_e)(\vec{k}_e \cdot
  \vec{k}_{\bar{\nu}})}{(E_e + m_e) E^2_e E_{\bar{\nu}}}\Big\}
\end{eqnarray}
and
\begin{eqnarray}\label{eq:B.5}
\hspace{-0.15in}&&\frac{d^5 \lambda^{(4)}_n(E_e, \vec{k}_e,
  \vec{k}_{\bar{\nu}}, \vec{\xi}_n, \vec{\xi}_e)}{dE_e d\Omega_e
  d\Omega_{\bar{\nu}}}\Big|_{(\rm N^2LO)} =  (1 + 3
g^2_A)\,\frac{G^2_F|V_{ud}|^2}{32\pi^5}\,(E_0 - E_e)^2 \,\sqrt{E^2_e -
  m^2_e}\, E_e\,F(E_e, Z = 1)\,\zeta(E_e) \nonumber\\
\hspace{-0.15in}&& \times \,3\, \frac{E_e}{m_N} \Big\{ -
\bar{K}_n(E_e)_{(\rm NLO)}\Big(\frac{(\vec{k}_e \cdot
  \vec{k}_{\bar{\nu}})^2}{E^2_e E^2_{\bar{\nu}}} - \frac{1}{3}\,
\frac{k^2_e}{E^2_e}\Big) - \bar{Q}_n(E_e)_{(\rm NLO)}
\frac{(\vec{\xi}_n \cdot \vec{k}_{\bar{\nu}})(\vec{k}_e \cdot
  \vec{k}_{\bar{\nu}})^2}{E^2_e E^3_{\bar{\nu}}} -
\bar{H}(E_e)_{(\rm NLO)}\nonumber\\
\hspace{-0.15in}&& \times \,\Big(\frac{(\vec{\xi}_e \cdot
  \vec{k}_{\bar{\nu}})(\vec{k}_e \cdot \vec{k}_{\bar{\nu}})}{E_e
  E^2_{\bar{\nu}}} - \frac{1}{3}\, \frac{\vec{\xi}_e \cdot
  \vec{k}_e}{E_e}\Big) - \bar{N}(E_e)_{(\rm NLO)} \frac{(\vec{\xi}_n
  \cdot \vec{\xi}_e)(\vec{k}_e \cdot \vec{k}_{\bar{\nu}})}{E_e
  E_{\bar{\nu}}} - \bar{Q}_e(E_e)_{(\rm NLO)} \, \frac{(\vec{\xi}_n
  \cdot \vec{k}_e)(\vec{k}_e \cdot \vec{\xi}_e)(\vec{k}_e \cdot
  \vec{k}_{\bar{\nu}})}{(E_e + m_e) E^2_e E_{\bar{\nu}}} \nonumber\\
\hspace{-0.15in}&& - \bar{K}_e(E_e)_{(\rm NLO)}\Big(\frac{(\vec{k}_e
  \cdot \vec{k}_{\bar{\nu}})^2}{E^2_e E^2_{\bar{\nu}}} - \frac{1}{3}\,
\frac{k^2_e}{E^2_e}\Big)\, \frac{\vec{\xi}_e \cdot \vec{k}_e}{E_e +
  m_e}\Big\} + 3\, \frac{E_e}{m_N}\Big(1
- \frac{\vec{k}_e \cdot \vec{k}_{\bar{\nu}}}{E_e
    E_{\bar{\nu}}}\Big)\frac{d^5 \lambda_n(E_e, \vec{k}_e,
    \vec{k}_{\bar{\nu}}, \vec{\xi}_n, \vec{\xi}_e)}{dE_e d\Omega_e
    d\Omega_{\bar{\nu}}}\Big|_{(\rm NLO)},
\end{eqnarray}
where the abbreviations ${\rm NLO}$ and ${\rm N^2LO}$ mean
``next-to-leading'' and ``next-to-next-to-leading'' order in the large
nucleon mass $m_N$ expansion. These terms are proportional to $1/m_N$
and $1/m^2_N$, respectively. The contribution to the electron--energy
and angular distribution, given by Eq.(\ref{eq:B.1}) was calculated in
\cite{Bilenky1959, Gudkov2006, Ivanov2013, Ivanov2017,
  Ivanov2018}. The contributions $\zeta(E_e)_{(\rm NLO)}$ and
$\bar{X}(E_e)_{(\rm NLO)}$ for $\bar{X} = \bar{a}, \bar{A}$ and so on
we take from \cite{Ivanov2013, Ivanov2017, Ivanov2018,
  Ivanov2019}. For completeness of the analysis of the corrections of
order $O(E^2_e/m^2_N)$ we adduce them in Appendix C.

\section*{Appendix C: Next-to-leading order corrections of order
  $O(E_e/m_N)$, caused by weak magnetism and proton recoil
  \cite{Ivanov2013, Ivanov2017, Ivanov2018, Ivanov2019}}
\renewcommand{\theequation}{C-\arabic{equation}}
\setcounter{equation}{0}

The next-to-leading order corrections $O(E_e/m_N)$ in the large
nucleon mass expansion, caused by weak magnetism and proton recoil, to
the neutron lifetime and correlation coefficients, which we denote as
$\zeta(E_e)_{(\rm NLO)}$ and $\bar{X}(E_e)_{(\rm NLO)}$, respectively,
for $\bar{X} = \bar{a}, \bar{A}$ and so on we take in the form
slightly corrected with respect to that calculated in
\cite{Ivanov2013, Ivanov2017, Ivanov2018, Ivanov2019}. They are given
by
\begin{eqnarray*}
\hspace{-0.3in}\zeta(E_e)_{(\rm NLO)} &=&\frac{1}{1 + 3
  g^2_A}\,\frac{E_0}{m_N}\, \Big[ - 2\,g_A\big(g_A + (\kappa +
  1)\big) + \big(10 g_A^2 + 4(\kappa + 1)\, g_A + 2\big)\,\frac{E_e}{E_0}
  \nonumber\\
\hspace{-0.3in}&-& 2 g_A\,\big(g_A + (\kappa +
1)\big)\,\frac{m_e}{E_0}\,\frac{m_e}{E_e}\Big],\nonumber\\
\hspace{-0.3in}\bar{a}(E_e)_{(\rm NLO)} &=& \frac{1}{1 + 3
  g^2_A}\, \frac{E_0}{m_N}\,\Big[2 g_A\,\big(g_A + (\kappa +
  1)\big) - 4 g_A \big(3 g_A + (\kappa +
  1)\big)\,\frac{E_e}{E_0}\Big],\nonumber\\
\hspace{-0.3in} \bar{A}(E_e)_{(\rm NLO)} &=&\frac{1}{1 + 3 g^2_A}\,
\frac{E_0}{m_N}\,\Big[\big(g^2_A + \kappa\, g_A - (\kappa + 1)\big) -
  \big(5 g^2_A + (3\kappa - 4)\,g_A - (\kappa +
  1)\big)\,\frac{E_e}{E_0}\Big],\nonumber\\
\hspace{-0.3in}\bar{B}(E_e)_{(\rm NLO)} &=& \frac{1}{1
  + 3 g^2_A}\, \frac{E_0}{m_N}\,\Big[- 2\,g_A\big(g_A + (\kappa +
  1)\big)  + \big(7 g^2_A + (3 \kappa + 8)\, g_A + (\kappa +
  1)\big)\,\frac{E_e}{E_0} \nonumber\\
\hspace{-0.3in}&-& \big(g^2_A + (\kappa + 2)\, g_A + (\kappa +
1)\big)\,\frac{m_e}{E_0}\,\frac{m_e}{E_e}\Big],\nonumber\\
\hspace{-0.3in} \bar{K}_n(E_e)_{(\rm NLO)} &=& \frac{1}{1 + 3
  g^2_A}\,\frac{E_0}{m_N}\,\Big(5 g^2_A + (\kappa - 4)\, g_A - (\kappa +
1)\Big)\,\frac{E_e}{E_0},\nonumber\\
\hspace{-0.3in} \bar{Q}_n(E_e)_{(\rm NLO)} &=& \frac{1}{1 + 3
  g^2_A}\,\frac{E_0}{m_N}\,\Big[ \Big(g^2_A + (\kappa + 2)\, g_A +
  (\kappa + 1)\Big) - \Big(7 g^2_A + (\kappa + 8)\,g_A + (\kappa +
  1)\Big)\, \frac{E_e}{E_0}\Big],\nonumber\\
\hspace{-0.3in}\bar{G}(E_e)_{(\rm NLO)} &=& \frac{1}{1 + 3
  g^2_A}\,\frac{E_0}{m_N}\,\Big[\big(2 g^2_A + 2 (\kappa + 1)\,
  g_A\big) - \big(10 g^2_A + 4(\kappa + 1)\,g_A +
  2\big)\, \frac{E_e}{E_0} \Big]\nonumber\\
\hspace{-0.3in} \bar{H}(E_e)_{(\rm NLO)} &=& \frac{1}{1 + 3
  g^2_A}\,\frac{E_0}{m_N}\,\frac{m_e}{E_e}\,\Big[ - 2\, g_A \big(g_A +
  (\kappa + 1)\big) + \big(4 g^2_A + 2 (\kappa + 1)\, g_A -
  2\Big)\,\frac{E_e}{E_0}\Big],\nonumber\\
\end{eqnarray*}
\begin{eqnarray}\label{eq:C.1}
\hspace{-0.3in} \bar{N}(E_e)_{(\rm NLO)} &=& \frac{1}{1 + 3
  g^2_A}\,\frac{E_0}{m_N}\,\frac{m_e}{E_e}\,\Big[-
  \Big(\frac{4}{3}\,g^2_A + \Big(\frac{4}{3} \kappa -
  \frac{1}{3}\Big)\,g_A - \frac{2}{3} (\kappa + 1)\Big)
  \nonumber\\
\hspace{-0.3in}&+& \Big(\frac{16}{3}\, g^2_A + \Big(\frac{4}{3} \kappa
- \frac{16}{3}\Big)\,g_A - \frac{2}{3} (\kappa + 1)\Big)\,
\frac{E_e}{E_0}\Big],\nonumber\\
\hspace{-0.3in}\bar{Q}_e(E_e)_{(\rm NLO)} &=&\frac{1}{1 + 3
  g^2_A}\,\frac{E_0}{m_N}\,\Big[- \Big(\frac{4}{3}\,g^2_A +
  \Big(\frac{4}{3} \kappa - \frac{1}{3}\Big)\,g_A - \frac{2}{3}
  (\kappa + 1)\Big) + \Big(2\,g^2_A + (2 \kappa +
  1)\,g_A\Big)\,\frac{m_e}{E_0} \nonumber\\
\hspace{-0.3in}&+& \Big(\frac{22}{3}\, g^2_A +
  \Big(\frac{10}{3}\kappa - \frac{10}{3}\Big)\,g_A - \frac{2}{3}
  (\kappa + 1)\Big)\, \frac{E_e}{E_0}\Big], \nonumber\\
\hspace{-0.3in} \bar{K}_e(E_e)_{(\rm NLO)} &=& \frac{1}{1 + 3
  g^2_A}\,\frac{E_0}{m_N}\, \Big[- 2 g_A \big(g_A + (\kappa + 1)\big)
  + \big(8 g^2_A + 2 (\kappa + 1) g_A + 2\big)\, \frac{m_e}{E_0} \nonumber\\
\hspace{-0.3in}&+& 4 g_A \big(3 g_A + (\kappa + 1)\big)\,
\frac{E_0}{E_e}\Big].
\end{eqnarray}
This corrections we give for the completeness of the analysis of
corrections of order $O(E^2_e/m^2_N)$, which we have carried out in
this paper. They should be used for the calculation of the
contributions $\bar{X}(E_e)_{(\rm LO)}\zeta^2(E_e)_{(\rm NLO)}$ and
$\bar{X}(E_e)_{(\rm NLO)}\zeta(E_e)_{(\rm NLO)}$ to the corrections
$X(E_e)_{(\rm N^2LO)}$ of order $O(E^2_e/m^2_N)$ (see
Eq.(\ref{eq:11})).

\end{document}